# Effect of doping, photodoping and bandgap variation on the performance of perovskite solar cells

*Basita Das*[A,B*], *Irene Aguilera*[A], *Uwe Rau*[A,B] and *Thomas Kirchartz*[A,C*]
[A]IEK5-Photovoltaik, Forschungszentrum Jülich, 52425 Jülich, Germany
[B]Faculty of Electrical Engineering and Information Technology, RWTH Aachen University, Mies-van-der-Rohe-Straße 15, 52074 Aachen, Germany
[C]Faculty of Engineering and CENIDE, University of Duisburg-Essen, Carl-Benz-Str. 199, 47057 Duisburg, Germany

*Abstract*

Most traditional semiconductor materials are based on the control of doping densities to create junctions and thereby functional and efficient electronic and optoelectronic devices. The technology development for halide perovskites had initially only rarely made use of the concept of electronic doping of the perovskite layer and instead employed a variety of different contact materials to create functionality. Only recently, intentional, or unintentional doping of the perovskite layer is more frequently invoked as an important factor explaining differences in photovoltaic or optoelectronic performance in certain devices. Here we use numerical simulations to study the influence of doping and photodoping on photoluminescence quantum yield as well as other device relevant metrics. We find that doping can improve the photoluminescence quantum yield by making radiative recombination faster. This effect can benefit or harm photovoltaic performance given that the improvement of photoluminescence quantum efficiency and open-circuit voltage is accompanied by a reduction of the diffusion length. This reduction will eventually lead to inefficient carrier collection at high doping densities. The photovoltaic performance might improve at an optimum doping density which depends on a range of factors such as the mobilities of the different layers and the ratio of the capture cross sections for electrons and holes.

# 1. Introduction

Halide perovskites (HP) exhibit excellent optoelectronic properties that manifest themselves in steep absorption onsets[1] and long carrier-diffusion lengths[2,3] that make sure that efficient light absorption and charge collection is possible in thin films of hundreds of nanometer thickness. The most peculiar property, however, is an excellent luminescence quantum efficiency[4–10] that substantially exceeds that of solution processed polycrystalline films made of other material families. It is this last property that has started extensive investigations into the defect tolerance[11–15] of the material class and has enabled highly efficient light emitting diodes as well as solar cells with open-circuit voltages[7,16–18] that so closely approach the radiative limit that they are only overcome by monocrystalline GaAs solar cells[19].

Halide perovskite devices so far have been made by employing a large library of contact materials[20–24] borrowed from organic solar cells, organic LEDs and dye-sensitized solar cells. The key concept of device making is based on the idea that the actual light absorbing or emitting layer is fairly intrinsic and that electron and hole injection or extraction has to be ensured by contacts with suitable workfunctions, electron affinities and ionization potentials[25]. The classical approach of doping the active layer as done in Si, III-Vs or other inorganic semiconductors has so far only rarely been pursued and currently there is no evidence that a functional perovskite-based pn-junction[26] solar cell or LED has been made. Nevertheless, unintentional doping via shallow intrinsic point defects is an important topic of research[27], triggered to a large degree by the evidence for mobile ionic charges[28,29] that lead to reversible transient effects and features like the *JV* curve hysteresis[29–32]. Mobile ions may appear in Frenkel pairs (i.e. for each positive ion there would be a negative ion) without actually doping the semiconductor by creating an excess of one type of charge. Therefore, it is not entirely obvious whether halide perovskites should be considered as doped semiconductors, which – given the typical thicknesses and permittivities – would be the case from doping densities of roughly $>10^{16}$ cm$^{-3}$ as found in Ref. 32. There is clear evidence in the most frequently studied composition methylammonium lead-iodide (MAPI) that the material behaves like an intrinsic semiconductor[33] with extremely low bulk charge densities ($<10^{12}$ cm$^{-3}$) measured in thick crystals[2]. Furthermore, in transient photoluminescence of thin films of MAPI[8], the signature of quadratic radiative recombination has been observed[8] down to (low) charge densities of $10^{14}$ cm$^{-3}$, suggesting doping densities that must be even lower. There are a variety of Mott-Schottky measurements[33] of lead-halide perovskites that show higher doping densities in the $10^{15}$ to $10^{17}$ cm$^{-3}$ range. However, the values approach those expected for an intrinsic semiconductor of the given thickness and are therefore no trustworthy evidence of doping. Among the more convincing evidence for doping is however a work that is based on a completely different approach not influenced by the intrinsic limitations of capacitance-based measurements[33]. The work from Feldmann et al.[9,10] reports on steady state photoluminescence (PL) and transient PL as well as transient absorption spectroscopy of a wide range of lead-halide perovskite absorber layers that are well suited for photovoltaics. The curious result of the study was that with the exception of the indeed quite intrinsic MAPI recipe also used in refs.[7,8], all other perovskite recipes were behaving as one would expect from a doped semiconductor. While MAPI has already quite good PL quantum yields, the Feldmann study reported even higher quantum yields for a set of different triple or quadruple cation perovskites, thereby raising the question whether the doping is an essential ingredient for achieving the highest PL quantum yields in this material class.

Here, we attempt to provide a theoretical basis for the findings of Feldmann *et. al.* (Refs. 9–10) and develop models to perform a critical assessment of the role of doping on PL quantum yield and photovoltaic device performance. One of the conclusions of Feldmann and colleagues was that the apparent doping originated from lateral band gap fluctuations that would cause local asymmetries between the concentrations of electrons and holes and that this level of disorder was actually promoting the observed performance gains. While it is an interesting thought that disorder could be beneficial for efficiency, the concept is in complete opposition to earlier findings in other photovoltaic communities where extensive theoretical work on the *negative* effect of band gap inhomogeneities was developed about 15 years ago[34–39]. Hence, in the current study we try to find answers to the following questions: (i) Can small lateral band gap fluctuations be the source of improved luminescence quantum efficiencies? (ii) What is the difference between doping or photodoping (by whatever mechanism) on the performance and functionality of photovoltaic devices? (iii) Is doping beneficial for solar cells and/or light emitting diodes and are there conditions that have to be met to benefit from doping?

In order to tackle the different questions, we use two approaches. First, we develop a simple analytical model to study the effect of lateral band gap variations and conclude that it cannot possibly be beneficial for device performance. In the second part of the paper, we explore doping by homogeneous concentrations of charged defects that are either always ionized (doping) or only ionized under illumination (photodoping) and describe the consequences of these two forms of doping on luminescence, charge transport and efficiency using numerical device modelling. We find that the open–circuit voltage improves with doping. However, the overall effect of doping on device performance depends on how doping affects the device at the maximum power point and is strongly correlated on the choice of non–radiative recombination coefficients[40–46]. If for instance, electron capture is slower than hole capture, moderate p-type doping can be particularly beneficial because it further reduces the slower of the two rates (the electron capture rate). Too high doping densities will however lead to a reduction in the diffusion length due to the shortening of the radiative lifetimes with doping and will eventually overcompensate all benefits from a reduction in Shockley–Read–Hall(SRH) recombination rates. Thus, for asymmetric capture coefficients there is typically a finite optimum doping density. Finally, we highlight and study the importance of transport through the electron and hole transport layers[47–50]. Depending on e.g. the charge carrier mobility in these layers, the concentration of electrons and holes at the maximum power point (MPP) varies substantially thereby modulating the amount of recombination losses for a given doping concentration, perovskite mobility and SRH lifetimes. Hence, the impact of doping or photodoping the absorber layer cannot be studied independently of the properties of the contact layers.

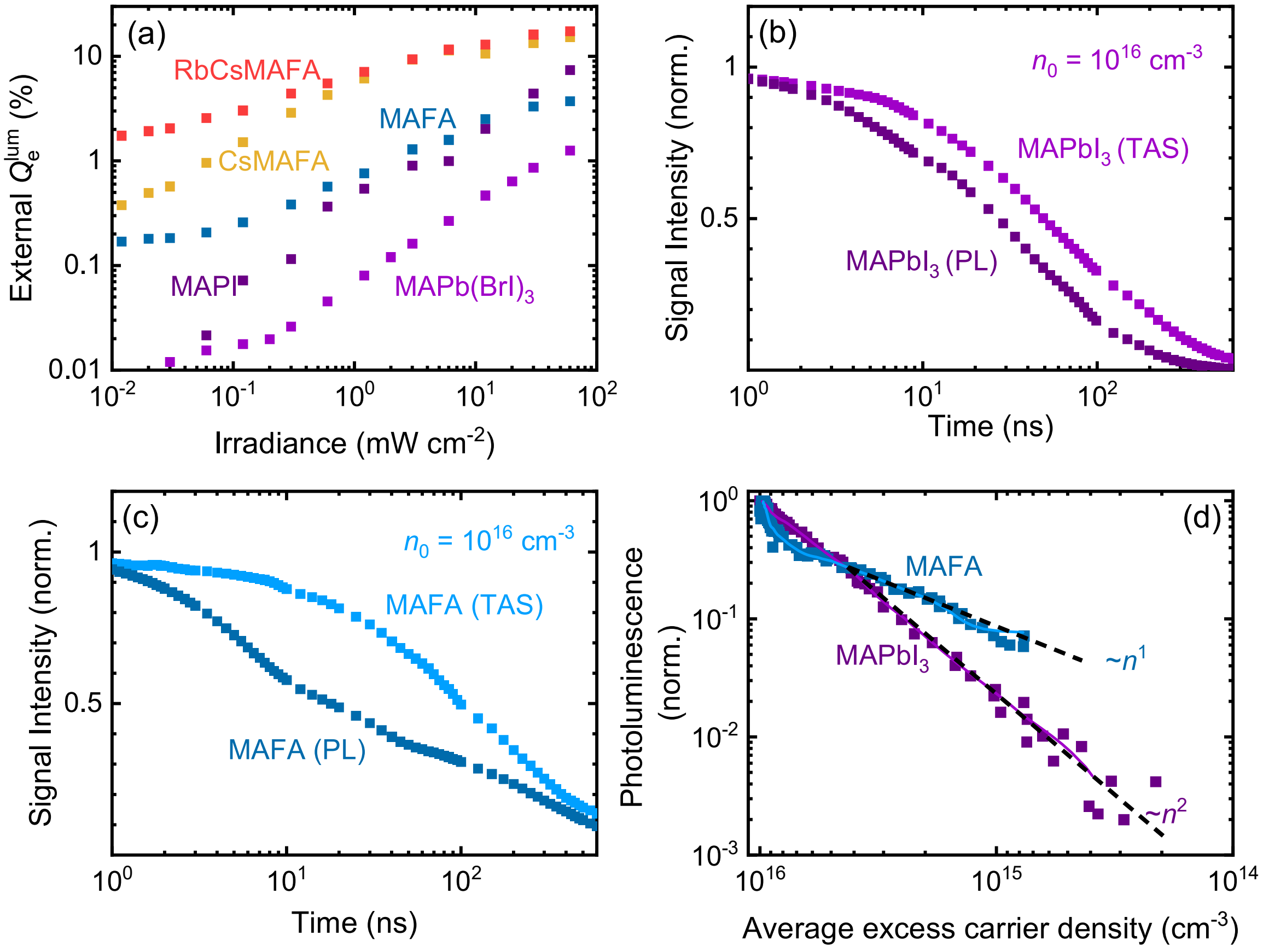

Figure. 1. Photoluminescence spectroscopy (PL) and transient absorption spectroscopy (TAS) measurements on MAPI and MAFA films with the data taken from ref [9]. (a) External photoluminescence quantum efficiency $Q_e^{\mathrm{lum}}$ of MAFA and MAPI films as a function of irradiance. (b) Normalized signal intensity of PL and TAS measurements on MAPI films. (c) Same measurements as in (b) but on MAFA thin films. (d) PL intensity plotted as function of the average excess carrier density obtained from TAS measurements in (b) and (c).

# 2. Role of spatial inhomogeneities

## 2.1. Revisiting the data of Feldmann et al in Refs. 9

Feldmann et. al. [9,10] observed that the external photoluminescence quantum efficiencies ($Q_e^{\mathrm{lum}}$) of mixed halide thin films vary depending on the composition of the thin films. They reported an increase from $Q_e^{\mathrm{lum}}$= 16.8% observed in methylammonium lead iodide (MAPI), to $Q_e^{\mathrm{lum}}$ = 22.6% in methylammonium-formamidinium based perovskites (MAFA) and subsequently to $Q_e^{\mathrm{lum}}$= 40.9% in potassium-passivated cesium methyl-ammonium formamidinium lead bromide-iodide (KCsMAFA). The quantum yields were measured under illumination intensities equivalent to one sun. They performed photoluminesece (PL) spectroscopy and transient absorption spectroscopy (TAS) to identify the recombination and total carrier population dynamics, respectively, on the MAPI, MAFA and KCsMAFA films. In Figure. 1(b and c) we plot the PL and TAS measurements for MAPI and MAFA films reported in [9]. The PL signal intensity indicates the amount of radiative recombination inside the device and the TAS signal intensity measures the average excess carrier concentration inside the device. If we assume that the density of states in conduction and valence band is identical, then the analysis of the ground state bleach feature of the TAS measurement would give us the average excess carrier concentration $\Delta n_{av} = (\Delta n + \Delta p) / 2$ in the sample[51]. Combining the two measurements gives us insight into the recombination dynamics as a function of photogenerated or excess charge carrier concentration inside the device[1]. In **Figure. 1**(d) we plot the PL intensity of the two films as a function of the excess carrier concentration $\Delta n$, obtained by multiplying the time dependent TAS signal intensity with the initial excess carrier density of $\Delta n_0 = 10^{16}\text{cm}^{-3}$, and time as an implicit variable. From the combined analysis of Figure. 1(d) it becomes apparent that in the MAPI thin films, the PL intensity decays with the square of the total carrier concentration whereas in the MAFA thin films it decays linearly with the total carrier concentration. The radiative recombination rate $R_{rad} = k_{rad}np$, in high injection scenario ($n = p = \Delta n$) simplifies to $R_{rad} = k_{rad}\Delta n^2$. Furthermore, when the thin film is acceptor doped with an acceptor concentration $N_A$, then the hole concentration remains constant $p = N_A$ and only the electron concentration changes with $\Delta n$. Hence, the radiative recombination rate, $R_{rad} = k_{rad}\Delta n N_A$, depends linearly on the excess carrier concentration $\Delta n$. Thus, if one of the carrier densities in the MAFA thin film is made constant by doping the PL signal becomes linearly proportional to $\Delta n$. In Figure. 1(a) we plot the fluence–dependent $Q_e^{\text{lum}}$ for MAPI and MAFA films and it is shown in the supporting information(SI)[9] that the increase in the fluence–dependent $Q_e^{\text{lum}}$ can be modelled as an influence of doping the thin film.

In order to explain the asymmetry between electron and hole concentrations, the authors of Ref. 9–10 suggest band gap inhomogeneities as a possible reason. This explanation is mostly based on a spatially resolved determination of the band gap derived from the PL peak measured with a confocal microscope[9]. Over several $\mu$m of lateral distance, the observed variations in the PL peak energy are about 20 meV between the different domains in a MAFA film. However, the reported change in bandgap energy of about 20 meV in the MAFA [$Ma_{0.2}Fa_{0.8}Pb(Br_{0.17}\ I_{0.83})_3$] films is much smaller as compared to that reported in other mixed–halide films[52–55]. The lateral bandgap variation in mixed halide film originates from halide–segregation which leads to separate narrow bandgap iodine rich and wide bandgap bromine rich phases in a perovskite films and can thus cause variations in band gap of hundreds of meV[52–55]. However, in the MAFA films studied here, the concentration of bromine is less than <20%, such that most of the film is dominated by narrow bandgap iodine rich phases, resulting in bandgaps in the range of 1.66eV – 1.68eV.

## 2.2. Modelling spatial inhomogeneities

To understand the impact of lateral bandgap variations on $Q_e^{\text{lum}}$ we developed a zero–dimensional(0D) model of a generic semiconductor thin film with two bandgaps as shown in **Figure. 2**(a). To implement the two–bandgap model we choose a valence band offset variation to make our model consistent with that of Feldmann's hypothesis[9]. We also chose flat–Fermi levels with the assumption that the lateral bandgap variations are local and are dimensionally smaller than the diffusion length of the carriers[37]. In our 0D model, the generic semiconductor thin film has a region with a low bandgap $E_{gL}$ and a region with a high bandgap $E_{gH}$ with a valence band offset $\Delta E_V$ between the two. The relative percentage of volume of each bandgap region in the thin film is determined by the volume fraction parameter $\beta$. When $\beta = 0$ or $\beta = 1$ then thin film has only one bandgap such that $E_g = E_{gL}$ or $E_g = E_{gH}$, respectively. When $0 < \beta < 1$, the volume of the high bandgap region is determined by $\beta$ and by $(1 - \beta)$ for that of the low bandgap region. Also, by design, the electron concentration $n_H$ and $n_L$ of the high and low bandgap region, respectively, are equal. Charge neutrality is maintained by adjusting the hole concentration $p_H$ and $p_L$ of the high and low bandgap region such that

$$n_{\mathrm{H}} = n_L = \beta p_{\mathrm{H}} + (1 - \beta) p_{\mathrm{L}} \quad (1)$$

holds. As the valence band offset $\Delta E_V$ increases between the two regions, $p_H$ decreases and $p_L$ increases as determined by $\beta$ and $\Delta E_V$ as defined by

$$p_{\mathrm{L}} = \frac{n_{\mathrm{H}}}{1 - \beta + [\beta \exp(-\Delta E_{\mathrm{V}}/k_{\mathrm{B}}T)]} \quad (2)$$

and

$$p_{\mathrm{H}} = p_{\mathrm{L}} \exp(-\Delta E_{\mathrm{V}}/k_{\mathrm{B}}T). \quad (3)$$

[1] Assuming the intrinsic carrier concentration $n_i$ is orders of magnitude smaller than the excess carrier concentration and total carrier concentration is approximately equal to excess carrier concentration.

The calculated carrier concentration is then used to determine various recombination rates. To make the model consistent, the radiative recombination coefficient $k_{rad}$ [cm$^3$/s] is determined for each different bandgap from the corresponding black body radiation spectrum as shown in eq. (S6 –S7) in the SI. The non–radiative recombination electron and hole coefficient $k_n$ and $k_p$, and the Auger coefficients for electrons and holes $c_n$ and $c_p$, respectively, are assumed from values available in literature[5]. For every type of recombination mechanism, the rates are calculated separately for the high and low bandgap and the effective rate is determined as a sum of the individual rates in each region, weighted by their volume fraction. For example, the radiative recombination rate for the low bandgap region $R_{radL}$ and that of the high bandgap region $R_{radH}$ will give the effective radiative recombination rate $R_{rad} = \beta \times R_{radH} + (1 - \beta) \times R_{radL}$. The internal photoluminescence quantum efficiency ($Q_{\mathrm{i}}^{\mathrm{lum}}$) is determined as a ratio between the radiative recombination rate and the total recombination rate in the film. The $Q_{\mathrm{e}}^{\mathrm{lum}}$ plotted in Figure. 2(b) is then calculated as[5]

$$Q_{\mathrm{e}}^{\mathrm{lum}} = \frac{p_{\mathrm{e}} Q_{\mathrm{i}}^{\mathrm{lum}}}{1 - \left(p_{\mathrm{r}} Q_{\mathrm{i}}^{\mathrm{lum}}\right)} \quad (4)$$

where photon emission probability $p_e$ and photon reabsorption probability $p_r$ are self consistently calculated from the absorptance and the black body spectrum for a specific bandgap. The mathematical construct of the model is given in table SI of the supporting information.

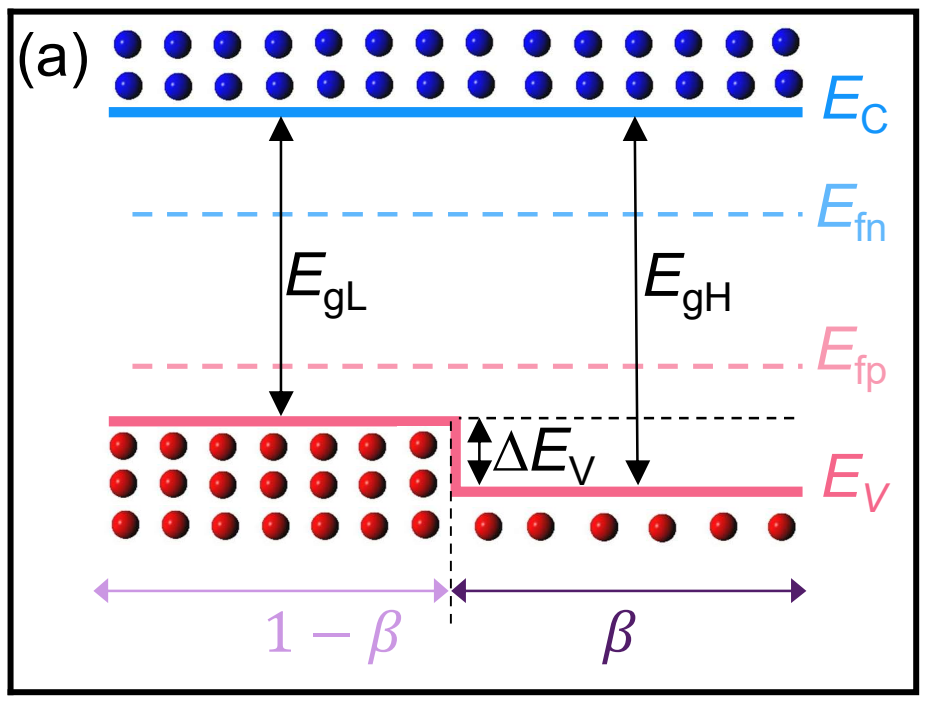

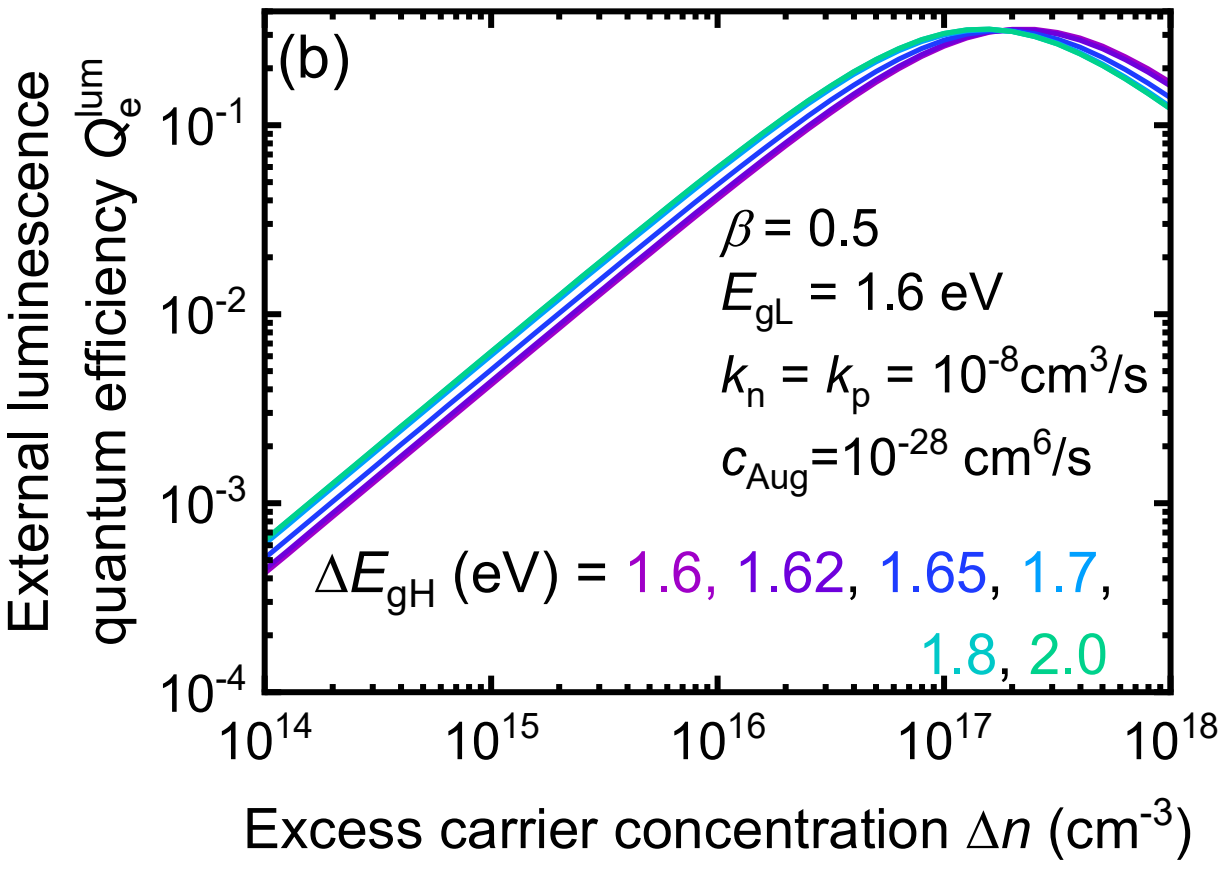

Figure. 2. Model to study the impact of lateral bandgap variations on PL yield and device performance. (a) Schematic representation of a 0D model implementing lateral bandgap variations in generic semiconductor thin films. The generic semiconductor film consists of two different regions identified by a lower bandgap $E_{gL}$ and a higher bandgap $E_{gH}$ with a valence band offset of $\Delta E_V$ between them and no conduction band offset for the sake of simplicity. The electron concentration in the two regions are equal as indicated by the constant distance between the conduction band $E_C$ and the electron quasi–Fermi level $E_{fn}$. However, the hole concentration is higher in the low bandgap region as compared to that in the high bandgap region as indicated by the distance between the valence band $E_V$ and hole quasi–Fermi level $E_{fp}$. The volume fraction $\beta$ determines the volume of the high bandgap region in the semiconductor thin film and (1 – $\beta$) determines the portion of the low bandgap region. (b) $Q_e^{lum}$ as a function of arbitrarily chosen excess carrier concentration and for various of high and low bandgap combinations. The low bandgap is kept fixed at $E_{gL}$=1.6 eV and $E_{gH}$ is varied.

From our zero–dimensional analysis we observe that $Q_e^{lum}$ varies very little with bandgap homogeneities [Figure. 2(b)] and is unlikely to show any variation for small bandgap inhomogeneities (0.01 to 0.02 eV) reported in Refs. 9. In a thin film with lateral bandgap variation, the recombination rates are supposed to decrease in the high bandgap region due to decrease in hole concentration $p_H$ and is supposed to increase in the low bandgap region due to increase in $p_L$. Subsequently when the weighted sum of the rates of the two regions is calculated, as shown in the example above, the effective rate will be mostly determined by the larger of the two individual rates. The recombination rates in the low bandgap region are orders of magnitude higher than that in the high bandgap region and thus the effective rates do not change much with bandgap inhomogeneities. As a result, very little variation is seen in $Q_e^{lum}$ and the general trend will be as shown in Figure. 2(b). Detailed analysis could be found in Figure. S1–S2 of SI.

Another question that needs answering in the present context is whether an increase in $Q_e^{lum}$ due to lateral bandgap variation ensures better photovoltaic performance. Rau and Werner[36] using a simple analytic model, quantifying the bandgap variations using standard deviations, emphasized that spatial variations of bandgap degrade the maximum achievable efficiency of the solar cell in the radiative limit and also affect non-radiative recombination. To understand the effect of lateral bandgap variation on solar cell

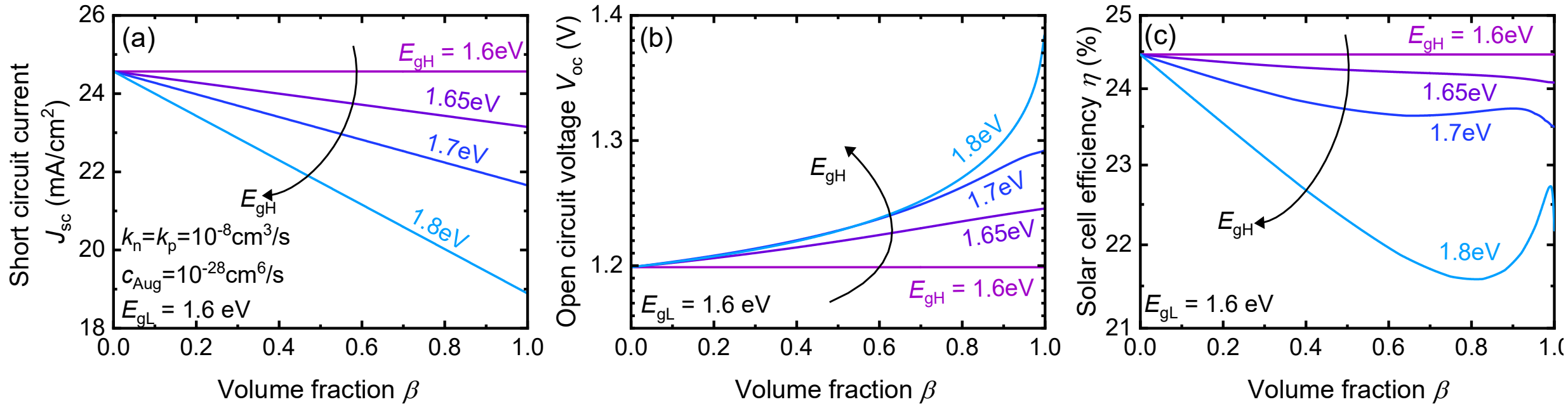

Figure. 3. Effect of lateral bandgap variation and volume fraction $\beta$ on short–circuit current $J_{sc}$, open–circuit voltage $V_{oc}$ and solar cell efficiency $\eta$. (a) $J_{sc}$ decreases as $E_{gH}$ and $\beta$ increases due to decrease of absorption at lower photon energies. (b) $V_{oc}$ increases as $E_{gH}$ and $\beta$ increases due to decrease non–radiative recombination at open–circuit condition. (c) With increasing volume fraction $\beta$ ,the solar cell efficiency $\eta$ decreases under the combined influence of $J_{sc}$ and $V_{oc}$ (except for high $E_{gH}$ and high $\beta$ where the increase in $J_{sc}$ overcompensates the loss in $V_{oc}$)

performance, we use our 0D model for various combinations of $E_{gL}$ and $E_{gH}$ and continuously varying $\beta$. In **Figure. 3** we plot the variation of short–circuit current $J_{sc}$, open circuit voltage $V_{oc}$ and solar cell efficiency $\eta$ as a function of volume fraction $\beta$ for different combinations of $E_{gL}$ and $E_{gH}$. In the 0D model, $J_{sc}$ is calculated by integrating the product of AM1.5G spectrum and absorptance over all photon energies (see equation S33 in SI). The absorption coefficient $\alpha$ [cm$^{-1}$] is a function of the bandgap of a material and the absorption edge (the energy from which a material starts absorbing incoming radiations) shifts to higher photon energies as the bandgap of the material increases and hence the absorptance *a* of the material also decreases for lower photon energies [see Figure. S3(a) in SI]. In our model the effective absorptance *a* of a two-bandgap thin film is given as a sum of the individual absorptances $a_L$ and $a_H$ of the lower and the higher bandgap region, respectively, weighted by their volume fraction. Thus, the absorptance for lower photon energy decreases [see Figure. S3(a) of SI] when the $E_{gH}$ of the higher bandgap region increases, and/or the volume of the higher bandgap region increases. As a result of less absorption at the lower photon energies, the short–circuit current decreases as shown in Figure. 3(a). Due to decrease in absorption at lower photon energies with increase in $E_{gH}$ and $\beta$, the overall generation rate $G$ [cm$^{-3}$s$^{-1}$] also decreases (see Figure. S3(b) of SI). Hence the equation $G = R_{tot}$ defining the open–circuit situation is attained at a lower value of the total recombination rate $R_{tot}$ with increase in $E_{gH}$ and $\beta$. The increase of the average band gap leads to a decrease of the equilibrium concentrations ($n_i^2$ goes down) and thereby to an increase of $np/n_i^2$ at open circuit, which leads to a higher open–circuit voltage since

$$V_{oc} = \frac{k_B T}{q} ln\left[\frac{np}{n_i^2}\right] \quad (5)$$

Where $k_B$ is the Boltzmann constant, $T$ is the temperature, $q$ is the elementary charge, $n$ and $p$ are the electron and hole concentration, respectively. The change in open–circuit voltage $V_{oc}$ agrees with that reported in literature as a result of bandgap variation due to phase segregation[52–55]. Mahesh et. al.[55] reported a decrease in open–circuit voltage with the increase in the ratio of the minority phases (narrow bandgap iodine rich phases) and we reproduce that same trend as we move from $\beta = 1$ to $\beta = 0$ (right to left) in Figure. 3(b). The decrease in $J_{sc}$ and increase in $V_{oc}$ with $E_{gH}$ and $\beta$ leads to decrease in solar cell efficiency $\eta$ as compared to the efficiency of the single bandgap thin film device with a lower bandgap $E_{gL}$ [see Figure. S3 (c–d)]. The efficiency might increase towards very high value of $E_{gH}$ and $\beta$ driven by the increase in open–circuit voltage. However, the single bandgap film $E_{gH} = E_{gL}$ [represented by the purple curve in Figure. 3(c)] gives the highest efficiency among all the combinations plotted in Figure. 3(c).

From the above analysis we infer that lateral bandgap variation in a thin film is not likely to improve the photovoltaic performance. Even though we would see an improvement in $V_{oc}$, the total charge generation would be limited by the increase in the higher bandgap value and /or with the increase in the volume fraction of the higher bandgap region and hence the short–circuit current, thereby eventually limiting the efficiency.

# 3. Results and discussion

## 3.1. Photodoping

In the following, we want to briefly discuss the terminology of doping and photodoping of a semiconductor using p-type semiconductors as an example. The more familiar concept of doping implies that there is a concentration of acceptor-like defects that is completely or at least partly ionized already without illumination or applied voltage (thermal and chemical equilibrium situation). Furthermore, the concentration of ionized acceptor-like defects has to greatly exceed the concentration of ionized donor-like defects for the semiconductor to be p-type. The acceptor-like defect will have to be close to the valence band such that in equilibrium, the Fermi level is always above the defect, thereby ensuring that it is ionized. To ensure the extended functionality, doping should not disappear during the operation of an (opto)electronic device and hence, also the electron and hole quasi-Fermi levels under normal operation conditions should both stay above the defect. This may impose even tighter constrains on the position of the defect level.

In contrast, the term photodoping refers to a situation where e.g. an acceptor-like defect is not ionized in equilibrium but becomes ionized only under illumination, i.e. when the quasi-Fermi levels are split. This requirement implies that the acceptor-like defect is above the equilibrium Fermi level in equilibrium (zero volt in the dark) and is between the two quasi-Fermi levels under illumination. Thus, its occupation changes with illumination. This puts a very stringent condition on the location of the defect, i.e., most likely the defect has to be above midgap but cannot be too close to the conduction band in order to not be always empty. Furthermore, photodoping also implies that the capture coefficients for hole and electron capture have to be extremely different for the defect to be fully (or mostly) ionized under illumination. The reason for this is that the acceptor like defect of our example has to be ionized, i.e. occupied by a sufficient density of electrons under illumination. However, the defect is acceptor like and therefore causes a high density of holes to exist in the valence band. To keep the high electron density in the defect the capture rate for electrons must be at least similar if not higher than the capture rate of holes despite the fact that there are many more holes in the valence band than electrons in the conduction band. To improve clarity, we will therefore present a more concrete example illustrating the requirements for and the influence of photodoping.

We assume an acceptor–like defect level with a defect density of $N_T = 10^{17}$cm$^{-3}$ inside the absorber of a MAPI solar cell at about 0.7 eV from the conduction band. The absorber layer of the device has a thickness of 300 nm and the hole transport layer (HTL) and the electron transport layer (ETL) layer are 20 nm each. The ETL and HTL are modelled with material parameters identifying generic organic transport layers and differing only in their values of electron affinity. We now have to assume extremely asymmetric capture coefficients to allow photodoping to be effective (we show some more examples of photodoping with less asymmetric capture coefficients in Fig. S5 of the SI). In this example, we assume that the acceptor–like defect level has an electron capture coefficient $k_n = 10^{-2}$ cm$^3$s$^{-1}$ and hole capture coefficient $k_p = 10^{-11}$ cm$^3$s$^{-1}$, i.e. nine orders of magnitude difference. Such highly asymmetric capture coefficients have been not reported in literature. However, we chose such a combination of capture coefficient (i) to make the photodoping comparable to doping using an identical concentration of acceptor dopants and (ii) to also make a point on how highly asymmetric the capture coefficients have to be for the defect to be able to fully ionize under illumination and photodope the device. **Figure. 4**(a) shows the band diagram of such a device at equilibrium. The grey dashed line in Figure. 4(a) represents the equilibrium–Fermi level $E_{fi}$ and the defect level, represented by the solid blue line, is mostly above the $E_{fi}$ and hence will be empty, i.e. not occupied by an electron. When the device is illuminated, charge carriers are photogenerated and the electron quasi–Fermi level $E_{fn}$ moves above the defect level as shown in Figure. 4(b). The occupation probability of a defect level[4], i.e. the probability of a defect being occupied by an electron is expressed as the ratio of the rates of the processes filling the defect level to that of all the processes taking place. When the defect level is positioned within the two quasi–Fermi levels, only the electron capture process and the hole capture process is relevant because, the electron and hole emission rates are much smaller in comparison. As a result, the defect occupation probability can be expressed as

$$f_{\mathrm{T}} = \frac{nk_{\mathrm{n}}}{nk_{\mathrm{n}} + pk_{\mathrm{p}}}, \tag{6}$$

where $nk_n$ is the electron capture rate and $pk_p$ is the hole capture rate. For $f_T \approx 1$, the electron capture process has to be much larger than the hole capture process such that $nk_n >> pk_p$ (see Figure. S4 in the SI). So, in a situation where $p >> n$, the electron capture coefficient and hole capture coefficient must be highly asymmetric such that $k_n >> k_p$, to allow faster electron capture. If the capture coefficients were any less asymmetric, then the electron filling and electron emptying rate of the defect will be comparable, and the defect will remain empty. If, however, the defect is below the hole Fermi level, the doping would persist also in the dark, i.e. the system would not be photodoped but simply doped in the conventional sense (see Fig S6 in the SI). Hence, photogenerated electrons are trapped by the defect from the conduction band, leaving an excess hole concentration in the valence band and effectively p–doping the entire absorber layer with a doping concentration of $N_A = N_T = 10^{17} cm^{-3}$. When we have less asymmetric capture coefficients, not all the defects will be ionized and hence the amount of doping will also be reduced (see Fig. S5 in SI).

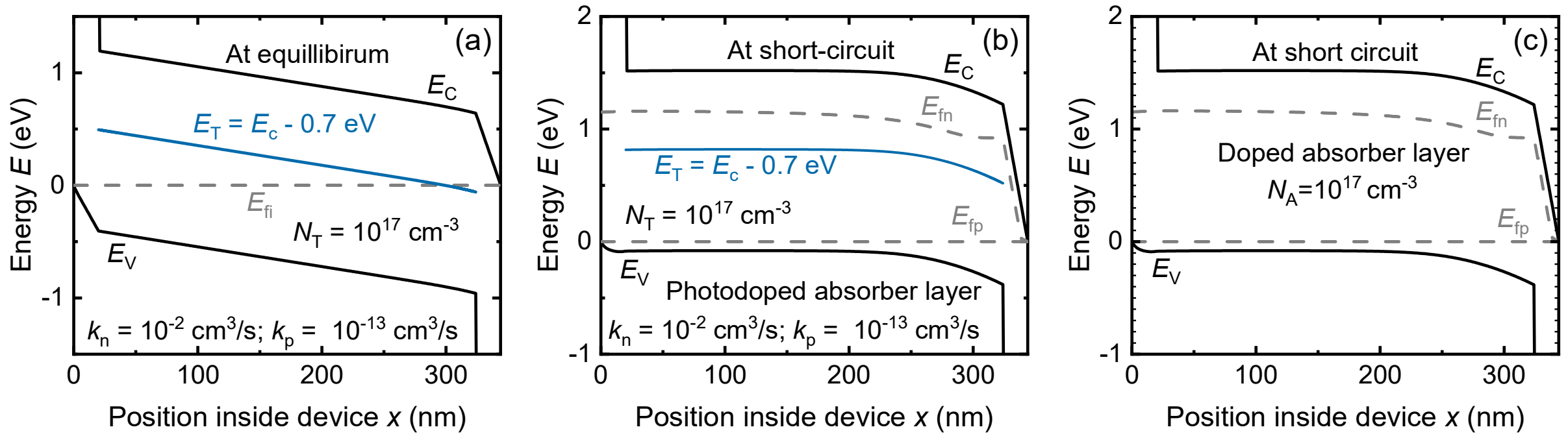

Figure. 4. Concept of photodoping and comparison with doping. (a) A generic perovskite solar cell, with a defect level 0.7 eV away from the conduction band and having asymmetric capture coefficients such that $k_n = 10^{-2}$ cm$^3$/s and $k_p = 10^{-13}$ cm$^3$/s. The device at equilibrium is not doped as indicated by the position of the intrinsic fermi level $E_{fi}$ at the middle of the bandgap. Also, the defect level is unoccupied by electrons since the $E_{fi}$ lies below the defect level. (b) The same device at short-circuit condition. The electron quasi–Fermi level $E_{fn}$ lies above the defect level and thus the defect level is occupied by electrons trapped from the conduction band. Hence there is an excess hole concentration in the valence band compared to the electrons in the conduction band and hence the device is effectively $p$–doped. (c) An acceptor doped device with identical device geometry to that in (a) but without a defect level at 0.7 eV away from the conduction band. The band diagrams resulting from photodoping (b) and from acceptor-doping (c) are virtually (practically) identical.

To draw a comparison between doping and photodoping, we consider another device identical in geometry to that of the photodoped device, but without the defect level with asymmetric capture coefficients and instead with a p–type absorber layer at a doping density of $N_A = 10^{17}$ cm$^{-3}$. Figure. 4(c) shows the band diagram of the device with the doped absorber and it is identical to the band diagram of the photodoped device in Figure. 4(b) at short–circuit. The similarity between doping and photodoping shown here ensures that the enhanced $Q_e^{lum}$ observed in [9] could be due to the presence of a highly asymmetric defect in the mixed halide films which gives the effect of doping by photodoping the film under illumination.

To summarize, for photodoping to occur in reality a series of very stringent conditions have to be fulfilled. The conditions being (i) the defect has to be above the equilibrium Fermi level at equilibrium condition, i.e., the defect will have to be between midgap and the conduction band, (ii) The occupation state of the defect has to change under illumination, i.e. the electron quasi–Fermi level has to move above the defect under illumination. This also restricts on how close the defect can be to the conduction band, (iii) Also if the defect is too close to conduction band then there is a fair chance of thermally activated electron emission, a process that will empty the defect of the electrons, and (iv) the electron capture coefficient of the defect has to be much larger than the hole capture coefficient such that the electron capture rate is orders of magnitude higher than the hole capture rate. So, it is more likely that enhanced $Q_{\mathrm{e}}^{\mathrm{lum}}$ observed in [9]is likely from doping and not photodoping.

Now that we have established that (i) the enhanced $Q_{\mathrm{e}}^{\mathrm{lum}}$ could be either from doping or photodoping (a less likely event) and (ii) the effect of doping and photodoping on the carrier concentration inside a device is identical, we will look into the question of whether an increase in $Q_{\mathrm{e}}^{\mathrm{lum}}$ via doping/photodoping of the absorber ensures an increase in the photovoltaic performance. We do so by looking at the effect of acceptor doping on the effective lifetime $\tau_{\mathrm{eff}}$, the internal luminescence quantum efficiency $Q_{\mathrm{i}}^{\mathrm{lum}}$ and the diffusion length $L_{\mathrm{n}}$ within the logic of our 0D model (see SI for equations). Acceptor doping a semiconductor layer with an acceptor doping density $N_{\mathrm{A}}$ gives a hole carrier density $p = N_{\mathrm{A}} + \Delta n \approx N_{\mathrm{A}}$ when $N_{\mathrm{A}} >> \Delta n$, and $p = N_{\mathrm{A}} + \Delta n \approx \Delta n$ when $N_{\mathrm{A}} << \Delta n$. When $N_{\mathrm{A}} << \Delta n$, we arrive at the high injection scenario such that $n = p = \Delta n$ whereas, when $N_{\mathrm{A}} >> \Delta n$, $n << p$. Recombination rates for different recombination mechanisms are functions of the respective recombination coefficients, such as $k_{\mathrm{rad}}$ for radiative recombination process, electron capture coefficient $k_{\mathrm{n}}$ and hole capture coefficient $k_{\mathrm{p}}$ for SRH recombination and $c_{\mathrm{n}}$ for Auger recombination involving two electrons and a hole and cp for Auger recombination involving one electron and two holes, and the electron and hole concentration in the layer as given below

$$R_{\mathrm{rad}} = k_{\mathrm{rad}}[np - n_i^2], \tag{7}$$

$$R_{\mathrm{SRH}} = N_T \frac{k_{\mathrm{n}} k_{\mathrm{p}}[np - n_i^2]}{nk_{\mathrm{n}} + pk_{\mathrm{p}} + e_{\mathrm{n}} + e_{\mathrm{p}}}, \tag{8}$$

$$R_{\mathrm{Aug}} = \left(c_{\mathrm{n}} n + c_{\mathrm{p}} p\right)[np - n_i^2]. \tag{9}$$

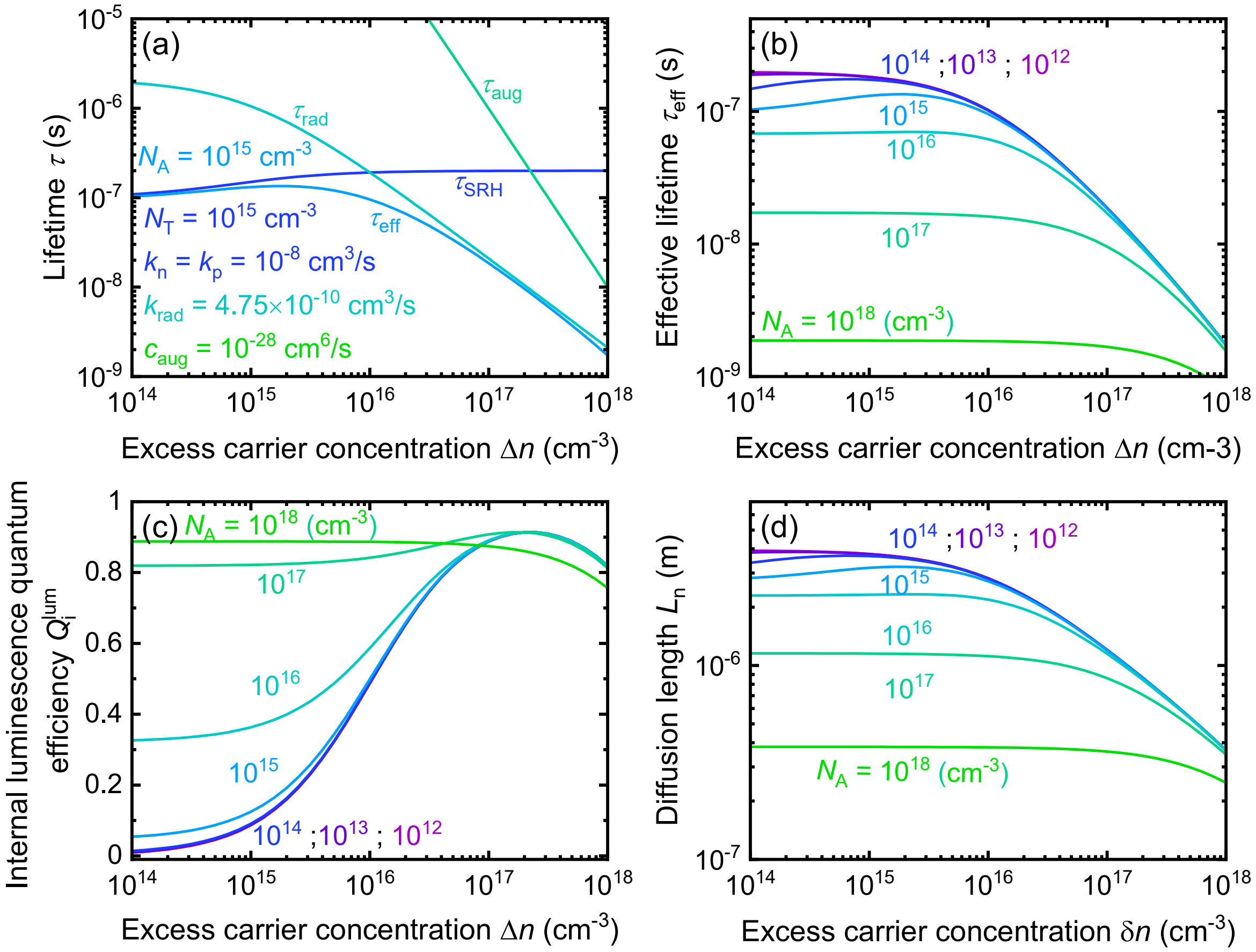

Figure. 5. Effect of doping on lifetime, internal luminescence quantum efficiency and diffusion length. (a) Variation of different recombination lifetime as a function of excess carrier concentration $\Delta n$ in the device at an acceptor doping density of $N_A = 10^{15}$ cm$^{-3}$. (b) Effective lifetime $\tau_{eff}$ for various doping densities and as a function of $\Delta n$. (c) Variation of $Q_i^{lum}$ with $N_A$ and $\Delta n$. $Q_i^{lum}$ increases with doping, thus having the possibility to positively affect the solar performance. (d) Variation of diffusion length $L_n$ with $N_A$ and $\Delta n$. $l_n$ decreases with doping thus with the possibility to adversely affect the solar performance.

The different recombination lifetimes are obtained from the corresponding recombination rates $R_x$ (x = SRH, rad, Aug) and the excess carrier concentration $\Delta n$ in the device according to $\tau_{x.} = \Delta n / R_x$. In **Figure. 5**(a) we plot the charge carrier lifetimes for radiative, Auger and SRH recombination for a fixed $N_A = 10^{15}$ cm$^{-3}$ and continuously varying $\Delta n$. We have assumed a defect density $N_T = 10^{15}$ cm$^{-3}$, Auger recombination coefficients $c_n = c_p = 5 \times 10^{-29}$ cm$^6$/s such that $c_{Aug} = c_n + c_p = 10^{-28}$ cm$^6$/s and calculated the recombination coefficient $k_{rad} = 4.75 \times 10^{-10}$ cm$^3$/s for $E_g = 1.6$ eV using the model described in Table S1 of SI. The choice of the electron and hole capture coefficients $k_n$ and $k_p$, respectively, is motivated by the calculations of Zhang *et. al*.[40] for iodine interstitial defect in methylammonium lead iodine perovskite. Zhang *et. al*. computed $k_n = 0.7 \times 10^{-8}$ cm$^3$/s and $k_p = 0.4 \times 10^{-4}$ cm$^3$/s. Thus, in Fig. 5–8, where we will try to understand how doping can affect the PSC performance, we treat the defect to have symmetric capture coefficients for simplicity. We assumed $k_n = k_p = 10^{-8}$ cm$^3$/s a value approximately equal to the electron capture coefficient $k_n$ predicted by Zhang et. al.[40] In the same plot we show the variation of effective lifetime given as the inverse sum of all the recombination lifetime $\tau_{\mathrm{eff}}^{-1} = \tau_{\mathrm{rad}}^{-1} + \tau_{\mathrm{SRH}}^{-1} + \tau_{\mathrm{Aug}}^{-1}$ and is limited by the fastest of all the recombination mechanism (smallest recombination lifetime).

At low excess charge carrier concentration $\Delta n$, the SRH recombination mechanism is the most dominant recombination mechanism and the effective lifetime $\tau_{\mathrm{eff}}$ [light blue curve in Figure. 5(a)] is limited by the SRH lifetime $\tau_{\mathrm{SRH}}$. As $\Delta n$ increases, the radiative recombination process becomes faster and the radiative recombination lifetime $\tau_{\mathrm{rad}}$ becomes comparable to $\tau_{\mathrm{SRH}}$ . Hence, $\tau_{\mathrm{eff}}$ becomes smaller (recombination becomes faster) under the influence of both mechanisms. As $\Delta n$ increases further the radiative recombination becomes the most dominating recombination mechanism and the effective lifetime $\tau_{\mathrm{eff}}$ is limited by $\tau_{\mathrm{rad}}$. Thus, in this regime the recombination becomes radiatively limited and the effective lifetime decreases with $\tau_{\mathrm{rad}}$. When $\Delta n$ is very high, the Auger recombination mechanism kicks in and the Auger lifetime decreases rapidly. However even in this regime $\tau_{\mathrm{rad}} \ll \tau_{\mathrm{Aug}}$ , thus influencing the effective lifetime only slightly.

In Figure. 5(b) we plotted the effective lifetime $\tau_{\mathrm{eff}}$ for various acceptor doping density as a function of $\Delta n$. We find that for our combination of capture coefficients, the recombination mechanism is mostly radiatively limited for acceptor doping concentration $N_{\mathrm{A}} \geq 10^{16}$ cm$^{-3}$. Also, for smaller doping concentrations $N_{\mathrm{A}} < 10^{16}$ cm$^{-3}$, the effective lifetime is radiatively limited for $\Delta n > 10^{16}$ cm$^{-3}$. As a result of this radiatively limited recombination mechanism for high doping densities or higher $\Delta n$, the internal luminescence quantum efficiency $Q_{\mathrm{i}}^{\mathrm{lum}}$ increases as shown in Figure. 5(c). The decrease in $Q_{\mathrm{i}}^{\mathrm{lum}}$ at very high values of $\Delta n$ and $N_{\mathrm{A}}$ is due to the increasing contribution of Auger recombination for very high values of $\Delta n$ and $N_{\mathrm{A}}$. So, from our analysis so far it appears that a higher doping concentration makes the recombination mechanism radiatively limited and hence might improve the open-circuit voltage of a solar cell made from such a material.

However, we must note that to improve photovoltaic performance, we not only have to reduce the fraction of non–radiative recombination but also must make sure that the photogenerated carriers are efficiently extracted. Diffusion length $L_{\mathrm{n}}$ is the parameter which quantifies the distance over which a photogenerated carrier can be transported by diffusion before it recombines. Diffusion length is a function of the mobility of the absorber layer as well as the effective lifetime of carrier in the same layer. It is calculated as $L_{\mathrm{n}} = \sqrt{(k_{\mathrm{B}}T\mu/q \times \tau_{\mathrm{eff}})}$, where $k_{\mathrm{B}}$ is the Boltzmann constant, $T$ is the temperature, $\mu$ is the mobility of the semiconductor material and $q$ is the elementary charge (see table SII in SI for material parameters used in the 0D model). The first term is often described as the diffusion coefficient $D$ [cm$^2$ s$^{-1}$]. The diffusion length, given its dependence on $\tau_{\mathrm{eff}}$, diminishes with increase in acceptor doping density $N_{\mathrm{A}}$ and excess carrier concentration $\Delta n$ as shown in Figure. 5(d).

Thus, doping of the absorber layer, has two contradictory effects, i.e., (i) the increase in $Q_{\mathrm{i}}^{\mathrm{lum}}$ due to increase in share of radiative recombination, which positively affects the photovoltaic performance, whereas (ii) decrease in $L_{\mathrm{n}}$ due to decrease in effective lifetime, which adversely affects the photovoltaic performance.

Inside a real device, whether doping will improve photovoltaic performance will depend on the interplay of the two effects of doping listed above. Besides, other factors like mobility of the transport layer, the asymmetric coefficients of recombination will also influence the impact of doping on photovoltaic performance. In the remainder of the paper we perform drift–diffusion simulations to analyze the impact of doping/photodoping, in combination with contact layer mobility and symmetric as well as asymmetric capture coefficients, on device performance.

## 3.2. Effect of doping/photodoping on perovskite solar cells

The drift–diffusion simulations performed in this section are done using the Advanced Semiconductor Analysis (ASA) software[56,57], an integrated opto-electronical tool developed by the Photovoltaic Materials and Devices group at TU Delft. For the device simulation we assume a generic p–i–n type perovskite solar cell (PSC) structure consisting of a MAPI layer sandwiched between an electron transport layer (ETL) and hole transport layer (HTL) before the cathode and anode layers, respectively. The two transport layers are symmetric, differing only in their value of electron affinity and are characterized by parameters identifying generic organic transport layers. The different values of electron affinity of the ETL and HTL are chosen to achieve carrier selectivity in the device, such that the ETL only allows electrons blocking holes and the HTL allows holes, blocking electrons. The different work functions of the cathode and anode metal establishes a built–in–voltage across the device. In PSC's the presence of mobile ions at the interfaces between the perovskite and transport layers screens the potential from the absorber layer, resulting in a field free absorber layer while most of the potential drops across the transport layers. However, ASA and many other semiconductor simulation tools do not explicitly consider (nonlocal) interfacial recombination between e.g. the conduction band of the ETL and the valence band of the absorber. To retain the ability to simulate interfacial recombination with non-zero band offsets at the perovskite- charge transport layer interfaces we include two very thin separate interface layers of 2 nm each. Each of the interfacial layers have the lower of the two adjacent conduction bands and the higher of the two valence bands. The mobility or permittivity are usually irrelevant in these layers because they are so thin. The defect density is set to a value such that it corresponds to the intended surface or interface recombination velocity (here 100 cm/s). Also, for simplicity, we assumed the workfunctions of the metal contacts such that the Schottky barrier height is zero at both the contacts.

To analyze the impact of doping/photodoping on PSC's we use the above-described geometry and vary (i) the doping concentration $N_A$ in the MAPI and the interface layers from $N_A = 10^{14} - 5 \times 10^{-17}$ cm$^{-3}$ and, (ii) the mobility $\mu_{TL}$ of the transport layer from $\mu_{TL} = 10^{-2} - 10^{-4}$ cm$^2$/Vs. We assume equal mobility for electron and holes in both ETL and HTL. Then we analyze the impact of varying $N_A$ and $\mu_{TL}$ on (i) open–circuit voltage $V_{oc}$, (ii) short–circuit current $J_{sc}$, and (iii) fill factor FF and solar cell efficiency $\eta$. We repeat the same analysis twice first with equal electron and hole capture coefficients $k_n = k_p = 10^{-8}$ cm$^3$/s and later with asymmetric capture coefficients identified for iodine interstitial defect by Zhang et.al.[40], where $k_n = 7 \times 10^{-9}$ cm$^3$/s and $k_p = 2 \times 10^{-5}$ cm$^3$/s. We assume a bulk donor defect density $N_T = 10^{15}$ cm$^3$/s, at 0.48 eV away from the conduction band, for both cases and an interface defect density $N_{int} = 10^{10}$ cm$^{-2}$ and $N_{int}$ = 1.4 × $10^{10}$ $cm^{-2}$ for symmetric and asymmetric capture coefficients, respectively, yielding a surface recombination velocity of $S \approx 100$ cm/s. For both sets of simulations, we assume a radiative recombination coefficient $k_{rad} = 4.75 \times 10^{-10}$ $cm^3$/s as calculated previously from our 0D model for a bandgap of 1.6 eV and Auger coefficients $c_{Aug} = 10^{-28}$ $cm^6$/s. All material and simulation parameters are listed in table SIII of SI. Also, the analysis given here is for a field free absorber case mimicking the screening of the potential by mobile ions at the interfaces.

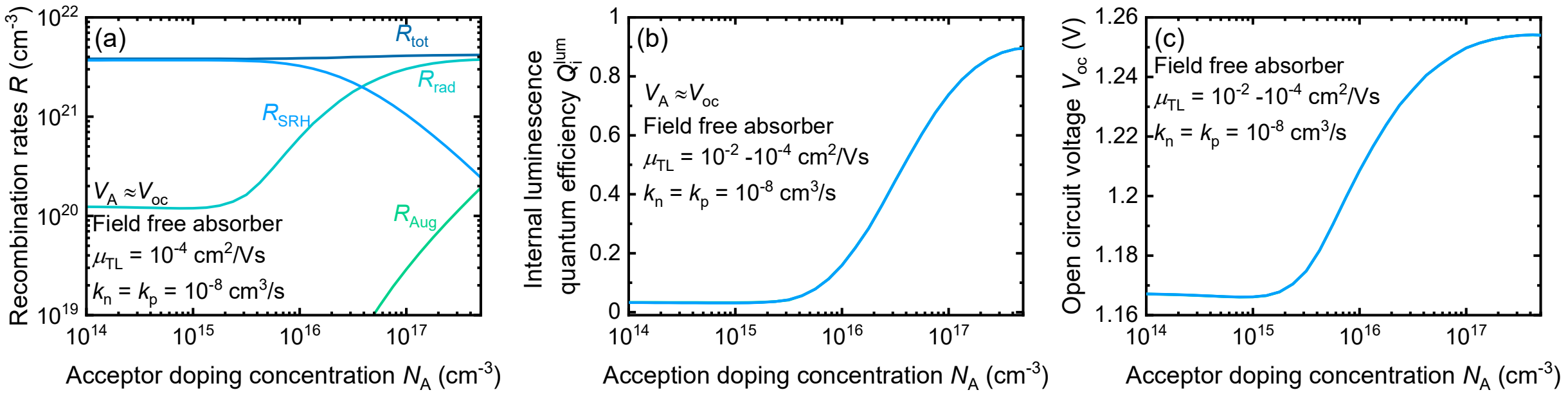

Figure. 6. Variation of recombination rate, internal luminescence quantum efficiency and open−circuit voltage as a function of acceptor doping density. (a) Variation of averaged SRH recombination rate $R_{SRH}$, radiative recombination rate $R_{rad}$, Auger recombination rate $R_{Aug}$ and total recombination rate $R_{tot}$ with $N_A$ at open−circuit condition for $\mu_{Tl} = 10^{-4}$ cm²/Vs. The $R_{SRH}$ decreases whereas $R_{rad}$ and $R_{Aug}$ increases at higher doping. $R_{tot}$ remains almost constant as combination of the three rates. (b) The internal quantum efficiency $Q_i^{lum}$ increases with $N_A$ because of the decrease in $R_{SRH}$ and increase in $R_{rad}$ at higher $N_A$ values. (c) The open−circuit voltage increases owing to the increase in the share of radiative recombination as $N_A$ increases. The mobilities of the transport layers do not affect the open circuit voltage because at open−circuit voltage irrespective of the value of the transport layer mobility we obtain flat−Fermi levels and no charge accumulation in the device, thus also keeping the recombination rates same for all mobilities.

### *3.2.1 Open−circuit voltage:*

In **Figure. 6** we plot the variation of the average recombination rates (integrated and averaged for all position inside the device between the two interfaces), the internal luminescence quantum efficiency (obtained from the averaged rates) at open circuit condition and that of the open−circuit voltage $V_{oc}$ as a function of the acceptor doping density and for multiple values of transport layer mobility $\mu_{TL}$. In Figure. 6(a) we plot the variation of the average SRH recombination rate $R_{SRH}$ for $\mu_{TL} = 10^{-4}$ cm²/Vs. The $R_{SRH}$ decreases for higher doping density, being limited by the electron capture rate. The electron capture rate $nk_n$ scales linearly with the electron concentration $n$ inside the device. The electron concentration decreases with increase in the hole concentration from increase in acceptor doping density. Hence, as the electron capture rate $nk_n$ decreases with increase in acceptor doping density, it becomes the rate limiting step in the SRH recombination and decreasing $R_{SRH}$. The radiative recombination rate $R_{rad}$ and the Auger recombination rate $R_{Aug}$ increases with the increase in hole concentration $p$ from increased acceptor doping. From the interplay of these three rates, the average total rate $R_{tot}$ remains almost constant, even though the share of the individual recombination mechanism changes.

As a result of the decreasing share of $R_{SRH}$ and increasing share of $R_{rad}$, the internal luminescence quantum efficiency $Q_i^{lum}$ and the open–circuit voltage $V_{oc}$ increase with doping concentration $N_A$. However, it is interesting to note that $Q_i^{lum}$ at open–circuit condition and $V_{oc}$ are not affected by the mobility of the transport layers. This is because there is no current flow through the transport layers at open–circuit. Hence, the recombination rates remain the same for all transport layer mobility and yields the same $Q_i^{lum}$ and open–circuit voltage $V_{oc}$ as shown in Figure. 6(b–c).

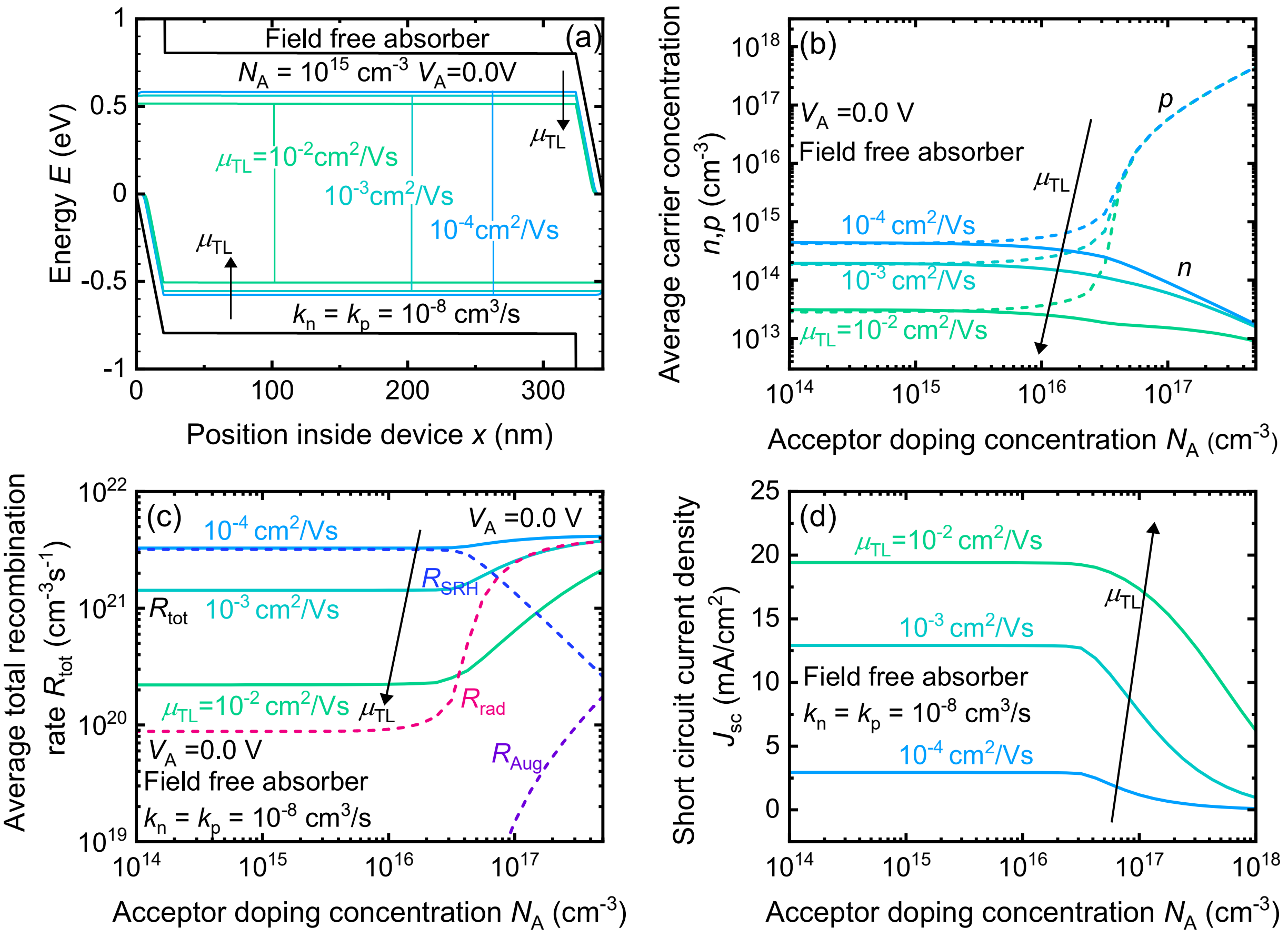

Figure. 7. The effect of doping on short–circuit current. (a) Band diagram of our generic device at short circuit for various values of transport layer mobility at $N_A$ = $10^{-15}$ $cm^{-3}$. As the transport layer mobility decreases, the electron and hole quasi–Fermi levels moves closer to their respective bands due to charge accumulation inside the device. (b) The variation of average electron *n* and hole *p* concentration inside the device with $N_A$. At low values of doping concentration, the average *n* and *p* values are identical. As the doping concentration increases, the hole concentration increases with it, whereas at the electron concentration decreases. When mobility of the transport layer decreases, the average electron and hole concentration increases inside the device due to higher charge accumulation inside the device. (c) The variation of different recombination rates. At low doping densities, the $R_{SRH}$ rate dominate the recombination landscape but as $N_A$ increases the radiative recombination rate increases and subsequently makes the total recombination radiatively limited. However, the total recombination still increases. Also, the recombination increases as transport layer mobility decreases owing to higher levels of electron and hole concentration inside the device due to charge accumulation. (d) The short–circuit current decreases at higher doping levels because of higher recombination. The short circuit current also decreases with decrease in transport layer mobility due to increase in recombination from higher charge accumulation.

### 3.2.2. Short−circuit current:

In this section we will analyze how the short−circuit current $J_{sc}$ varies with $N_A$ and $\mu_{TL}$. In **Figure. 7**(a) we plot the band diagram of our model device for $N_A$ = $10^{15}$ $cm^{-3}$ and $\mu_{TL}$ = $10^{-2}$ – $10^{-4}$ $cm^2/Vs$. The average electron and hole concentration inside the device are plotted as a function of $N_A$ for the three values of $\mu_{TL}$ in Figure. 7(b). The average electron and hole concentrations plotted in panel (b) were initially symmetric for lower doping concentration because the doping concentration is smaller than the excess photogenerated carrier concentration at 1 sun illumination. However, at higher doping densities, the doping concentration is higher than the photogenerated carrier density and hence the hole concentration increases with the acceptor doping density while the electron concentration decreases. The carrier concentration inside the device is also impacted by the mobility of the transport layers. Panel (b) shows that the average carrier concentration increases in the device as we decrease the mobility of the transport layer. This finding is corroborated by the movement of the electron and hole quasi−Fermi levels closer to their respective band edges in Figure. 7(a). The lower the mobility of the transport layer, the higher is the amount of charge accumulation inside the device due to poor charge extraction by the ETL and HTL.

In Figure. 7(c) we plot the average total recombination rates at short circuit. We observe that for low doping densities the recombination is limited by the $R_{SRH}$ while for higher doping densities, the recombination increases due to an increase in radiative recombination $R_{rad}$ = $k_{rad}(\Delta n_{sc}(\Delta n_{sc}+N_A))$ once $N_A$ exceeds the photogenerated charge density $\Delta n_{sc}$ at short circuit. As explained in the previous section, the $R_{SRH}$ recombination decreases for higher doping densities being limited by the electron capture rate and the $R_{rad}$ and $R_{Aug}$ increases under the influence of the increasing hole concentration. The dashed lines show the increase and decrease of these average rates at short−circuit for $\mu_{TL}$ = $10^{-4}$ $cm^2/Vs$ and forms a guide to the eye for understanding the trend of average $R_{tot}$. As the mobility of the transport layer decreases, the $R_{tot}$ increases for lower doping densities under the influence of the $R_{SRH}$, which increases due the accumulation of charges inside the device.

In Figure. 7(d) we plot the change in short−circuit current $J_{sc}$ with $N_A$ and $\mu_{TL}$. $J_{sc}$ decreases for higher doping densities due to higher total recombination. $J_{sc}$ also decreases as we decrease the transport layer mobility due to increased total recombination from increase in non−radiative recombination.

### 3.2.3 Fill−factor and efficiency:

**Figure. 8**(a) shows that the fill factor *FF* decreases even though the open circuit voltage increases [Figure. 6(c)] with higher doping densities. The FF decreases with increase in doping density due to increased non–radiative recombination away from the open–circuit condition as can be seen from the current–voltage curve presented later in **Figure. 10**.

So, with the increase in doping (i) $V_{oc}$ increases due to increased share of radiative recombination at open–circuit condition, (ii) $J_{sc}$ decreases due to increase in overall recombination at short–circuit and, (iii) FF decreases due to increased non–radiative recombination away from the open–circuit condition. Thus, as a cumulative impact of doping on the three parameters listed here, the efficiency $\eta$(%) of the solar cell also decreases as shown in Figure. 8(b). The FF and $\eta$(%) also decrease with the decrease in transport layer mobility because more and more charge accumulation takes place and the SRH recombination increases.

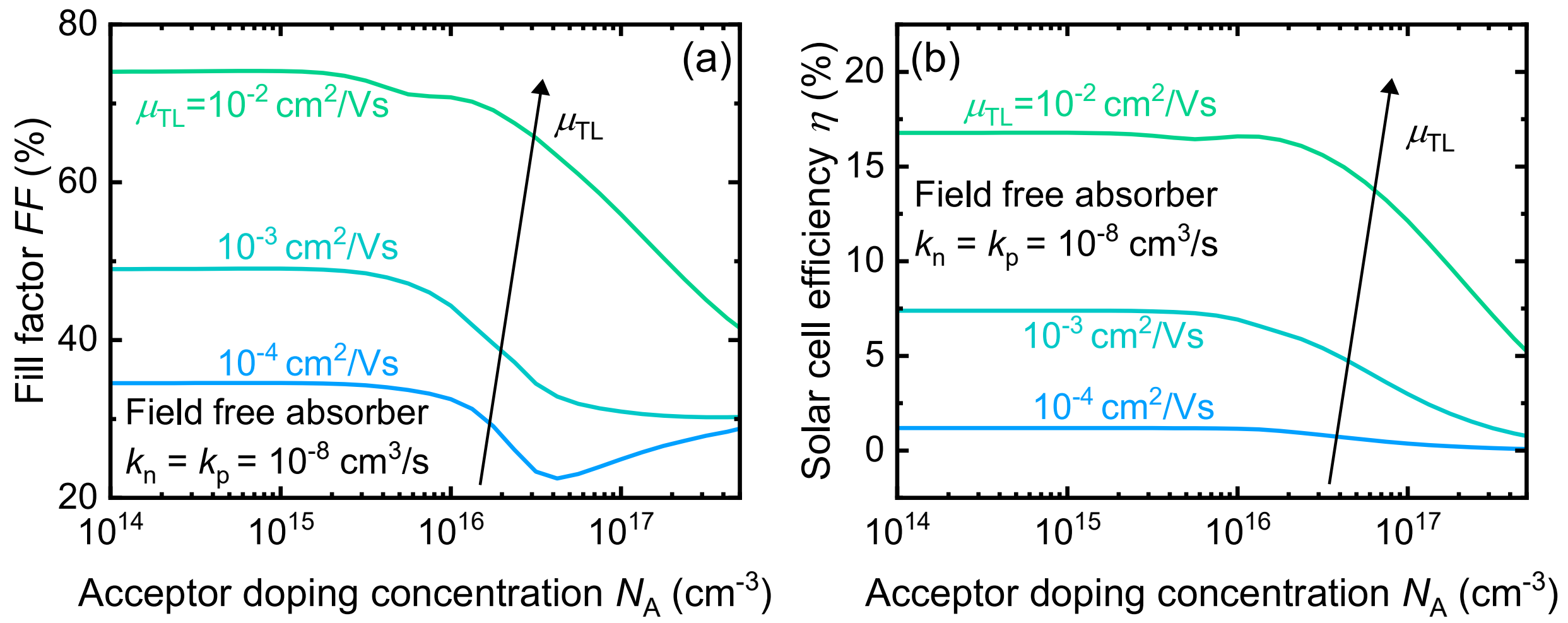

Figure. 8. Variation of fill factor and efficiency with doping density. (a) Fill factor FF decrease owing to decrease in short–circuit current and higher levels of recombination away from the open–circuit voltage with increase of doping density. FF also decreases with decrease in mobility of the transport layer due to increase in recombination from charge accumulation. (b) The efficiency of the solar cell decreases with increase in $N_A$ due to decrease in $J_{sc}$ and FF with increase in $N_A$ and transport layer mobility.

At low doping densities the average electron $n$ and, hole $p$ concentration inside the device were equal [Figure. 7(b)] and mostly constant. The SRH recombination rate also follows the trend of $n$ and $p$ in the low doping density regime. However, when the doping increases beyond $10^{15}$ cm$^{-3}$, and $n$ starts slowly decreasing and $p$ starts slowly increasing, the $R_{SRH}$ still remains same as before, until around $N_A > 3 \times 10^{16}$ cm$^{-3}$. So, in this regime of intermediate doping density ($10^{15} < N_A < 3 \times 10^{16}$ cm$^{-3}$), the $R_{SRH}$ do not follow the downhill trend of $n$. This is because in this regime of intermediate $N_A$ values, the values of $n$ and $p$ are not quite equal, but their difference is small. When this is combined with our assumption of the symmetric capture coefficients, it leads to a situation where the electron and hole capture rate are comparable to each other and the SRH rate is given by the combination of both capture rates and there is not a single rate-limiting step. However, had this been otherwise, such that the electron capture coefficient $k_n$ was much smaller than $k_p$, then the $R_{SRH}$ would have decreased with the decrease of electron concentration in the regime of intermediate doping density ($10^{15} < N_A < 3 \times 10^{16}$ cm$^{-3}$). Also, in this regime of intermediate doping density, the radiative recombination remains low, and hence there is a possibility that the total recombination rate would decrease, and we might be able to see an improvement in the solar cell performance. In the following section we will analyze such a case where capture coefficients are asymmetric.

### *3.2.4 Asymmetric capture coefficients:*

Defect levels in materials are either more likely to capture electrons or holes, but usually not equally likely to capture both carriers[4,40,41,58,59]. This property is quantified by asymmetric capture coefficients. Zhang et al.[40] have shown that iodine interstitial defects, which are a prime candidate for the most dominant recombination center in methylammonium lead iodide (MAPI) perovskites, have high hole capture coefficients and low electron capture coefficients. The origin of this asymmetry lies in the anharmonic shape of the potential energy surfaces [4,40]. In our analysis, we use the capture coefficients $k_n = 7\times10^{-9}$ cm$^3$/s and $k_p = 2\times10^{-5}$ cm$^3$/s calculated by Zhang et. al[40] from first principle for an iodine interstitial defect about 0.48 eV from the conduction band edge. We replace the donor defect level with symmetric capture coefficients in our generic device, described at the beginning of section. B, by one with asymmetric capture coefficients and keep everything else identical. We also adjust the defect density at the interfaces to keep the surface recombination velocity at $S$ = 100 cm/s.

When we introduce the asymmetric coefficients, we expect that the changes in trends of various quantities would be subtle in comparison to the case we analyzed in the previous section. The trend of internal luminescence quantum efficiency and open–circuit voltage is the same as in the symmetric case and they increase as we increase our doping density as shown in **Figure. 9**(a–b). Again, the mobility of the transport layers does not impact the $Q_\mathrm{i}^\mathrm{lum}$ and the $V_\mathrm{oc}$ owing to the flat Femi levels at open–circuit independently of the chosen value of $\mu_\mathrm{TL}$.

In the case of asymmetric coefficients ($k_n << k_p$), the electron capture rate $nk_n$ is the rate limiting step for the SRH recombination. As a result of this we see a decrease in the total recombination rates around intermediate doping density regime ($10^{15} < N_A < 3 \times 10^{16}$ cm$^{-3}$) as predicted in the last section. The decrease in the total recombination rate is due to the decrease in the SRH recombination rate as shown by the short–dashed lines in Figure. 9(c). The short–circuit current $J_\mathrm{sc}$ increases in this intermediate doping density regime as is shown in Figure. 9(d) due to the decrease in total recombination. Consequently, in Figure. 9(e), the fill factor FF increases in the range $10^{15} < N_A < 3 \times 10^{16}$ cm$^{-3}$ because of a decrease in non–radiative recombination away from open circuit[4]. As a result of the increase in $V_\mathrm{oc}$, $J_\mathrm{sc}$ and FF, the efficiency $\eta$ also increases in the range $10^{15} < N_A < 3 \times 10^{16}$ cm$^{-3}$. Thus, it is indeed true that when we have asymmetric capture coefficients, moderate levels of doping can help in reducing recombination around the maximum power point and can lead to improved photovoltaic performance. In our recent publication[4] we showed that SRH recombination through iodine interstitial defects with asymmetric capture coefficients can be limited by acceptor doping of the transport layer. Doping the hole transport layer leads to excess hole carriers in the MAPI absorber layer which reduces the electron capture rate and hence the SRH recombination rate through the defect level. Doping the hole transport layer thus allowed us indirectly modulating the carrier concentration inside the absorber layer to our advantage, without having to go into the complexity of doping the perovskite layer.

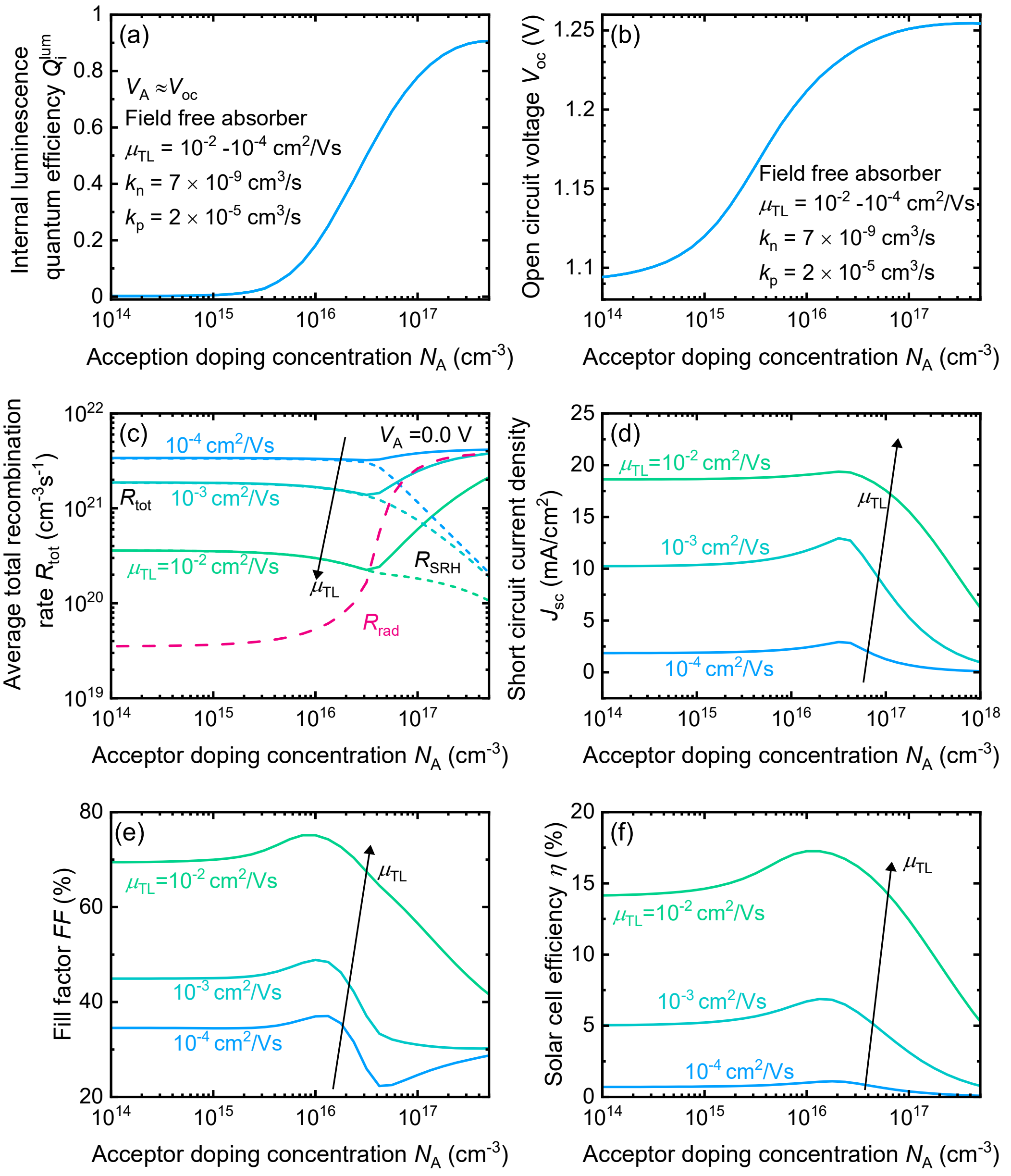

Figure. 9. The variation of $V_{oc}$, $J_{sc}$, FF and $\eta$(%) with $N_A$ in a PSC with a defect characterized by asymmetric capture coefficients. (a) Variation of $Q_i^{lum}$ at open circuit condition. $Q_i^{lum}$ increase in $N_A$ from increase in radiative recombination at increase in $N_A$ at open–circuit. (b) $V_{oc}$ also increases with $N_A$ and is radiatively limited. (c) Variation in recombination rates at short–circuit. The total recombination rate decreases at intermediate levels of doping $10^{15} < N_A < 3 \times 10^{16}$ cm$^{-3}$ due to decrease in $R_{SRH}$. The $R_{SRH}$ decreases having been limited by the electron capture rate $nk_n$. The electron capture rate decreases linearly with the decrease in electron concentration with increase in $N_A$. (d) As a result of the decrease in total recombination rate $R_{tot}$ in the doping range of $10^{15} < N_A < 3 \times 10^{16}$ cm$^{-3}$, the short–circuit current increase in this range of doping. (e) The FF increase due to increase in $J_{sc}$ as well as decrease in SRH recombination away from the open–circuit at intermediate levels of doping $10^{15} < N_A < 3 \times 10^{16}$ cm$^{-3}$. (f) The efficiency increases for intermediate levels of doping $10^{15} < N_A < 3 \times 10^{16}$ cm$^{-3}$ due to increase in $V_{oc}$ , $J_{sc}$ and $FF$. The trend of all the parameters with decrease in transport layer mobility remains identical to that in the symmetric capture coefficient case discussed in Figures 7 and 8.

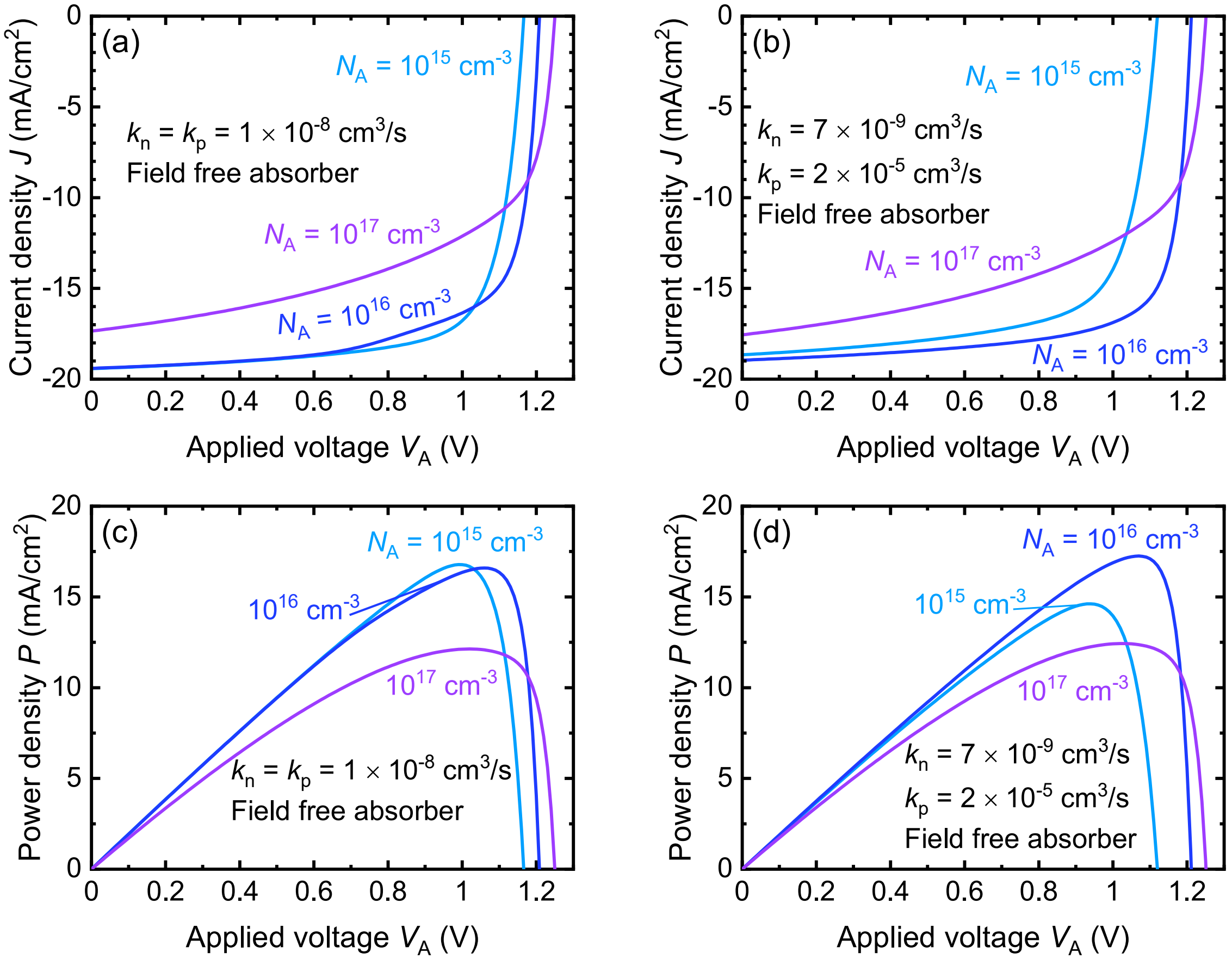

Figure. 10. Comparison of current–voltage and power–voltage curves at different doping densities for the symmetric and asymmetric capture coefficient case. (a) *JV* curves at $N_A$ = $10^{15}$,$10^{16}$ and, $10^{17}$ cm$^{-3}$ and $\mu_{TL}$=$10^{-2}$ cm$^2$/Vs from the device with defects characterized by symmetric capture coefficients. SRH recombination around maximum power point increases with higher doping densities. (b) *JV* curves from the device with defects characterized by asymmetric capture coefficients. The SRH rate decrease at maximum power point for $N_A$ = $10^{16}$ cm$^{-3}$. (c) PV curve from the device with symmetric capture coefficients. The maximum power output decreases with increase in $N_A$ (d) PV curve from the device with asymmetric capture coefficients. The maximum power output is obtained for $N_A$ = $10^{16}$ cm$^{-3}$.

In Figure. 10 we plot the current–voltage and power–voltage curves for both the symmetric as well as the asymmetric case to draw the comparison between the two cases. In the symmetric capture coefficient case, doping only decreases the power output [Figure. 10(a and c)] whereas in the asymmetric capture coefficient case, doping increases the power output by reducing SRH recombination away from open circuit as seen in Figure. 10(b and d). Thus, doping the absorber with the aim of reducing the carrier density associated with the rate-limiting step may indeed improve performance.

# 4. Conclusion

It has recently been proposed by Feldmann and co-workers[9,10] that photodoping due to small lateral band gap fluctuations in halide perovskites contributes to the superior luminescence quantum efficiencies in a variety of different multi-cation compositions. These high luminescence quantum efficiencies are of paramount importance both for applications in light emitting devices but also for photovoltaics. This is due to the relation between open-circuit voltage and luminescence quantum efficiency that links efficient luminescence to high $V_{oc}$ values relative to thermodynamic limits of $V_{oc}$.

These findings by Feldmann et al. raise several fundamental questions that ask for additional scrutiny: (i) Can small lateral band gap fluctuations be the source of improved luminescence quantum efficiencies? (ii) What is the difference between doping in photodoping (by whatever mechanism) on the performance and functionality of photovoltaic devices? (iii) Is doping beneficial for solar cells and/or light emitting diodes and are there conditions that have to be met to benefit from doping?

To answer these questions, we use different modelling approaches that include analytical modelling of lateral bandgap variation as well as drift–diffusion simulations of generic perovskite solar cells. The effort leads to the following conclusions: (i) From our analytical modelling it was confirmed that small lateral bandgap variation is not the likely source of photodoping but instead it arises from the presence of defects with highly asymmetric capture coefficients which preferentially captures one type of carriers creating an imbalance of carrier concentration between the two bands. (ii) We explained the origins of photodoping and a series of stringent conditions that need to be met for a device to be photodoped. We compared the impact of doping and photodoping on the carrier concentration inside the device and found that the two processes have identical impact on the carrier concentration inside the device under illumination Given the series of conditions for successful photodoping, we conclude that photodoping is unlikely to be the reason of enhanced luminescence quantum efficiencies observed by Feldmann et. al.[9,10]. (iii) From our solar cell device simulations, we found strong correlation between the effects of doping on device efficiency and the capture kinetics of the dominant recombination levels in the device. To optimize the performance of both solar cells as well as LEDs via doping, it is important to have knowledge of the capture coefficients of the defect level to make an informed choice on the type as well as amount of doping that will ensure the reduction in the share of non–radiative recombination. In this context, we highlight recent successes in calculating capture coefficients for both electrons and holes in lead-halide perovskites[40–42] which are likely to be highly beneficial for the understanding and controlled optimization of perovskite-based optoelectronic devices. Furthermore, for use as solar cells, the mobility of the transport layers plays an important role in ensuring that sufficient carriers get extracted before recombining.

# 5. Conflicts of interest

There is no conflict of interest.

# 6. Acknowledgements

All the authors acknowledge Dr. Feldmann and co–authors for kindly making available all the data from Ref. 9. B.D. acknowledges the HITEC graduate school at Forschungszentrum Jülich for support from a PhD. I.A. acknowledges funding from the European Commission Horizon 2020 project No. 824158 (“EoCoE-II”). T.K. acknowledges the Helmholtz Association for funding via the PEROSEED project.

# 7. Author contributions

IA,UR and TK supervised the project. BD performed the simulations and prepared the manuscript.

# Supporting information

# Effect of doping, photodoping and bandgap variation on the performance of perovskite solar cells.

*Basita Das*[A,B], *Irene Aguilera*[A], *Uwe Rau*[A,B] and *Thomas Kirchartz*[A,C]
[A]IEK5-Photovoltaik, Forschungszentrum Jülich, 52425 Jülich, Germany
[B]Faculty of Electrical Engineering and Information Technology, RWTH Aachen University, Mies-van-der-Rohe-Straße 15, 52074 Aachen, Germany
[C]Faculty of Engineering and CENIDE, University of Duisburg-Essen, Carl-Benz-Str. 199, 47057 Duisburg, Germany

## 1. Analytical model for lateral bandgap variation

Table SI. Mathematical construct of the 0D model

| Parameters | Equation | |
|---|---|---|
| Absorption coefficient [cm$^{-1}$] | $\alpha_{\mathrm{H/L}} = \alpha_0\sqrt{\frac{E-E_{gH/L}}{KT}}$ for $E > (E_{gH/L} + E_U)/2$ | (S1) |
| | $\alpha_{\mathrm{H/L}} = \alpha_0 \exp\left(\frac{E-E_{\mathrm{gH/L}}}{E_{\mathrm{U}}}\right)\sqrt{\frac{E_{\mathrm{U}}}{2\exp(1)kT}}$ for $E < (E_{\mathrm{gH/L}} + E_{\mathrm{U}})/2$ | (S2) |
| Urbach energy (eV) | $E_{\mathrm{U}}$ | |
| Bandgap (eV) | $E_{\mathrm{gH/L}}$; H: high bandgap ; L : Low bandgap | |
| Energy of the incoming radiation (eV) | $E$ | |
| | $\alpha_0 = 2 \times 10^4$ cm$^{-1}$ | |
| Absorptance | $a_{\mathrm{H/L}}(\mathrm{E}) = 1 - \exp(-\alpha_{\mathrm{H/L}}d)$ | (S3) |
| Thickness of the thin film [nm] | $d$ | |
| Effective absorptance | $a(E) = \beta a_{\mathrm{H}} + (1-\beta)a_{\mathrm{L}}$ | (S4) |
| Generation rate [cm$^{-3}$s$^{-1}$] | $G = \frac{\int a(E) \times \phi_{sun}(E)dE}{d}$ | (S5) |
| Black body spectrum [cm$^{-2}$s$^{-1}$eV$^{-1}$] | $\phi_{\mathrm{BB(H/L)}}(E) = \frac{2\pi a_{\mathrm{H/L}}(E)E^2}{h^3c^2}\exp\left(-\frac{E}{k_{\mathrm{B}}T}\right)$ | (S6) |
| Radiative recombination coefficient [cm$^3$s$^{-1}$] | $k_{\mathrm{rad(H/L)}} = \frac{\int_0^\infty 4\alpha_{\mathrm{H/L}}\eta_{\mathrm{r}}^2\phi_{\mathrm{bb(H/L)}}(E)dE}{n_i^2}$ | (S7) |
| Intrinsic carrier concentration [cm$^{-3}$] | $n_{\mathrm{iH/L}}^2 = \sqrt{N_{\mathrm{c}}N_{\mathrm{v}}}\exp\left[\frac{-E_{\mathrm{gH/L}}}{2k_BT}\right]$ | (S8) |
| Refractive index | $\eta_r$ | |
| Emission probability | $p_{\mathrm{e(H/L)}}=\frac{\int_0^\infty \alpha_{\mathrm{H/L}}\phi_{\mathrm{bb(H/L)}}(E)dE}{\int_0^\infty 4\alpha_{\mathrm{H/L}}\mathrm{d}\eta_r^2\phi_{\mathrm{bb(H/L)}}dE}$ | (S9) |
| Reabsorption probability | $p_{\mathrm{r(H/L)}}=1-p_{\mathrm{e(H/L)}}$ | (S10) |
| Effective emission probability | $p_{\mathrm{e}}=\beta p_{\mathrm{eH}}+(1-\beta)p_{\mathrm{eL}}$ | (S11) |
| Effective reabsorption probability | $p_r=\beta p_{\mathrm{rH}}+(1-\beta)p_{\mathrm{rL}}$ | (S12) |
| Excess carrier concentration[cm$^{-3}$] | $\delta n=n_{\mathrm{H}}=n_{\mathrm{L}}$(arbitrarily chosen) | |
| Condition of charge neutrality | $\delta\mathrm{n}=\beta p_{\mathrm{H}}+(1-\beta)p_{\mathrm{L}}$ | (S13) |
| Hole conc, in $E_{\mathrm{gL}}$ region [cm$^{-3}$] | $p_{\mathrm{L}}=\frac{\delta n}{\beta\exp\left[-\frac{\Delta\mathrm{E_v}}{k_{\mathrm{B}}T}\right]+1-\beta}$ | (S14) |
| Hole conc, in $E_{\mathrm{gH}}$ region [cm$^{-3}$] | $p_{\mathrm{H}}=p_{\mathrm{L}}\exp\left[-\frac{\Delta E_{\mathrm{v}}}{K_B T}\right]$ | (S15) |
| Radiative recombination rate [cm$^{-3}$s$^{-1}$] | $R_{\mathrm{rad(H/L)}}=k_{\mathrm{rad(H/L)}}\left(\delta n\, p_{\mathrm{H/L}}-n_{\mathrm{i(H/L)}}^2\right)$ | (S16) |
| Effective radiative recombination rate [ cm$^{-3}$s$^{-1}$] | $R_{\mathrm{rad}}=\beta R_{\mathrm{radH}}+(1-\beta)R_{\mathrm{radL}}$ | (S17) |
| Radiative lifetime [s] | $\tau_{\mathrm{rad}}=\delta\mathrm{n}/\mathrm{R_{rad}}$ | (S18) |
| SRH recombination rate [ cm$^{-3}$s$^{-1}$] | $R_{\mathrm{SRH(H/L)}}=N_{\mathrm{T}}\frac{k_{\mathrm{n}}k_{\mathrm{p}}\left(\delta n\, p_{\mathrm{H/L}}-n_{\mathrm{iH/L}}^2\right)}{nk_{\mathrm{n}}+pk_{\mathrm{p}}+e_{\mathrm{n(H/L)}}+e_{\mathrm{p(H/L)}}}$ | (S19) |
| Effective SRH recombination rate [ cm$^{-3}$s$^{-1}$] | $R_{SRH}=\beta R_{\mathrm{SRH(H)}}+(1-\beta)R_{\mathrm{SRH(L)}}$ | (S20) |
| SRH lifetime [s] | $\tau_{\mathrm{rad}}=\delta n/R_{SRH}$ | (S21) |
| Capture coefficients [cm$^3$s$^{-1}$] | $k_{\mathrm{n}}=k_{\mathrm{p}}$ | |
| Electron emission coefficients [s$^{-1}$] | $e_{\mathrm{n(H/L)}}=k_{\mathrm{n}}N_{\mathrm{C}}\,exp[-E_{\mathrm{T}}/k_{\mathrm{B}}T]$ | (S22) |
| Hole emission coefficients [s$^{-1}$] | $e_{\mathrm{p(H/L)}}=k_p N_V\,exp\left[\left(E_{\mathrm{T}}-E_{\mathrm{g(H/L)}}\right)/k_{\mathrm{B}}T\right]$ | (S23) |
| Auger recombination rate [ cm$^{-3}$s–1] | $R_{\mathrm{Aug(H/L)}}=\left(c_n\delta n+c_p p_{\mathrm{H/L}}\right)\left(\delta n\, p_{\mathrm{H/L}}-n_{\mathrm{iH/L}}^2\right)$ | (S24) |
| Auger coefficients | $c_n=c_p$ | |
| Effective Auger recombination rate [s$^{-1}$] | $R_{Aug}=\beta R_{\mathrm{Aug(H)}}+(1-\beta)R_{\mathrm{Aug(L)}}$ | (S25) |
| Auger lifetime [s] | $\tau_{\mathrm{Aug}}=\delta n/R_{\mathrm{Aug}}$ | (S26) |
| Total recombination [s$^{-1}$] | $R_{\mathrm{tot}}=R_{\mathrm{rad}}+R_{\mathrm{SRH}}+R_{\mathrm{Aug}}$ | (S27) |
| Effective lifetime [s] | $\tau_{eff}=\delta n/R_{tot}$ | (S28) |
| Diffusion length [cm] | $L_n=\sqrt{\frac{\mu k_B T}{q}\tau_{eff}}$ | (S29) |
| Internal Photoluminescence quantum efficiency | $Q_{\mathrm{i}}^{\mathrm{lum}}=R_{\mathrm{rad}}/\mathrm{R_{tot}}$ | (S30) |
| External photoluminescence quantum efficiency | $Q_e^{\mathrm{lum}}=\left\|p_{\mathrm{e}}Q_{\mathrm{i}}^{\mathrm{lum}}/\left(1-p_{\mathrm{r}}Q_{\mathrm{i}}^{\mathrm{lum}}\right)\right\|$ | (S31) |
| Voltage [V] | $V=\frac{k_{\mathrm{B}}T}{q}\ln\left[\frac{\delta n p_{\mathrm{H}}}{n_{\mathrm{iH}}^2}\right]=\frac{k_{\mathrm{B}}T}{q}\ln\left[\frac{\delta n p_L}{n_{\mathrm{iL}}^2}\right]$ | (S32) |
| Open circuit voltage [V] | $\mathrm{V_{oc}}=\frac{k_{\mathrm{B}}T}{q}\ln\left[\frac{\delta n_{\mathrm{(oc)}}p_{\mathrm{H(oc)}}}{n_{\mathrm{iH}}^2}\right]=\frac{k_{\mathrm{B}}T}{q}\ln\left[\frac{\delta n_{(oc)}p_{L(oc)}}{n_{\mathrm{iL}}^2}\right]$ | (S33) |
| Short circuit current [mAcm$^{-2}$] | $J_{\mathrm{sc}}=\mathrm{q}\int_0^\infty a(E)\phi_{\mathrm{sun}}(E)dE$ | (S34) |
| Recombination current [mAcm$^{-2}$] | $J_{\mathrm{rec}}=qdR_{\mathrm{tot}}$ | (S35) |
| Current [mAcm$^{-2}$] | $J=J_{\mathrm{rec}}-J_{\mathrm{sc}}$ | (S36) |
| Power [mAcm$^{-2}$] | $P=-JV$ | (S37) |
| Efficiency [%] | $\eta=\frac{\max(P)}{100}\times 100$ | (S38) |
| | | |

# 2. Effect of lateral bandgap variation

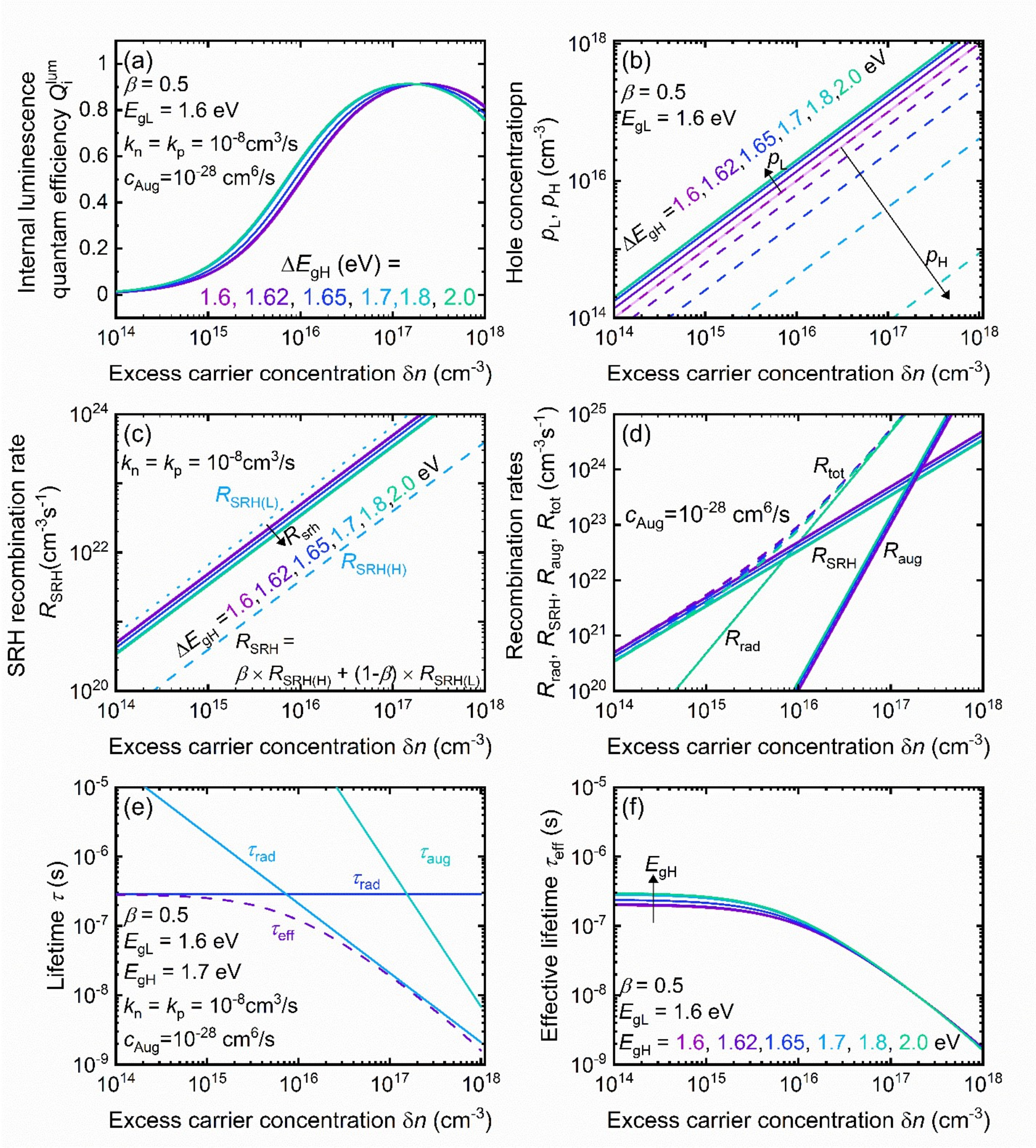

Figure S11. Effect of lateral bandgap variation on hole concentration, recombination rates, and internal luminescence quantum efficiency for various combinations of lower and higher bandgaps. (a) Internal luminescence quantum efficiency $Q_i^{lum}$ as a function of arbitrarily chosen excess carrier concentration $\delta n$ and for various of high and low bandgap combinations. The $Q_i^{lum}$ changes very little with bandgap variation as explained in panel (b–f). The corresponding $Q_e^{lum}$ is plotted in figure 2(b) in the main paper. (b) The variation of the hole concentration in high and low bandgap region is shown as a function of $\delta n$. In the high bandgap region, $p_H$ decreases with increase in higher bandgap value whereas in the low bandgap region $p_L$ increases. When $\beta = 0.5$ the change in $p_H$ and $p_L$ is quantitively identical but is different when $\beta \neq 0.5$. (c) The SRH recombination rate in the two regions, identified by two different bandgaps, also changes under the effect of $p_H$ and $p_L$. The dashed blue line shows the $R_{SRH}$ in the higher bandgap region and the dotted blue line shows the same in the low bandgap region for the combination $E_{gH} = 1.7$ eV and $E_{gL} = 1.6$ eV. The solid lines represent the effective $R_{SRH}$ for various combinations of $E_{gH}$ and $E_{gL}$ and show very little variation with change in the $E_{gH}$. The effective $R_{SRH}$ is dominated by the SRH rate in the lower bandgap region. (d) The same is true for all the other recombination mechanisms. Effective recombination rates of SRH, radiative as well as Auger mechanism for different combinations of $E_{gH}$ and $E_{gL}$ are shown. None of the effective rates have much variation with change in $E_{gH}$. As a result, the total effective recombination rate as shown by the dashed line does not vary much with change in $E_{gH}$ either. Hence, the $Q_i^{lum}$ only increases

when the total effective rate is dominated by the effective radiative rate at high $\delta n$ but do not show any change for low $\delta n$ (e) Variation of different recombination lifetime when $E_{gH}$ = 1.7 eV and $E_{gL}$ = 1.6 eV. The effective recombination lifetime is limited by the fastest of all the processes. (f) Effective recombination lifetime for different combinations of $E_{gH}$ and $E_{gL}$.

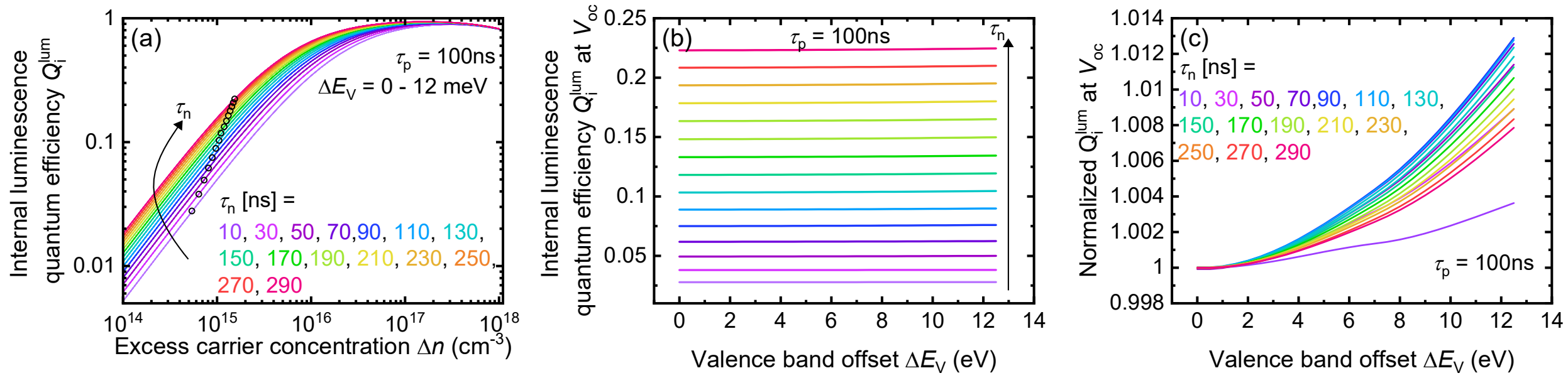

Figure S2: Effect of lateral bandgap variation on internal luminescence quantum efficiency $Q_{\mathrm{i}}^{\mathrm{lum}}$ for asymmetric electron and hole lifetimes and volume faction of high bandgap region $\beta = 0.2$. (a) $Q_{\mathrm{i}}^{\mathrm{lum}}$ is plotted as a function of the excess carrier concentration $\Delta n$ in the device for various combinations of electron and hole lifetimes $\tau_n$ and $\tau_p$ respectively for different values of valence band offset $\Delta E_V$. The $E_{gL}$ is kept fixed at 1.6 eV and the $E_{gH}$ is varied to realize the valence band offset between the two regions on the device. The lines for different values of $\Delta E Q_{\mathrm{i}}^{\mathrm{lum}}$ at $V_{oc}$ for all values of $\Delta E_V$ overlaps since there is very little variation for this small of a change in lateral bandgap. However, $Q_{\mathrm{i}}^{\mathrm{lum}}$ increases as the electron lifetime slows down. The black circles show the $Q_{\mathrm{i}}^{\mathrm{lum}}$ at open circuit voltage ($V_{oc}$) for each combination of $\tau_n$ and $\tau_p$. (b) $Q_{\mathrm{i}}^{\mathrm{lum}}$ at $V_{oc}$ for different $\tau_n$ and $\tau_p$ combination as a function of valence band offset $\Delta E_V$ and shows almost negligible variation along the x–axis. (c) To make the negligibly small variation more explicit, we plot $Q_{\mathrm{i}}^{\mathrm{lum}}$ at $V_{oc}$ for all values of $\Delta E_V$ normalized by the $Q_{\mathrm{i}}^{\mathrm{lum}}$ at $V_{oc}$ when $\Delta E_V = 0$eV.

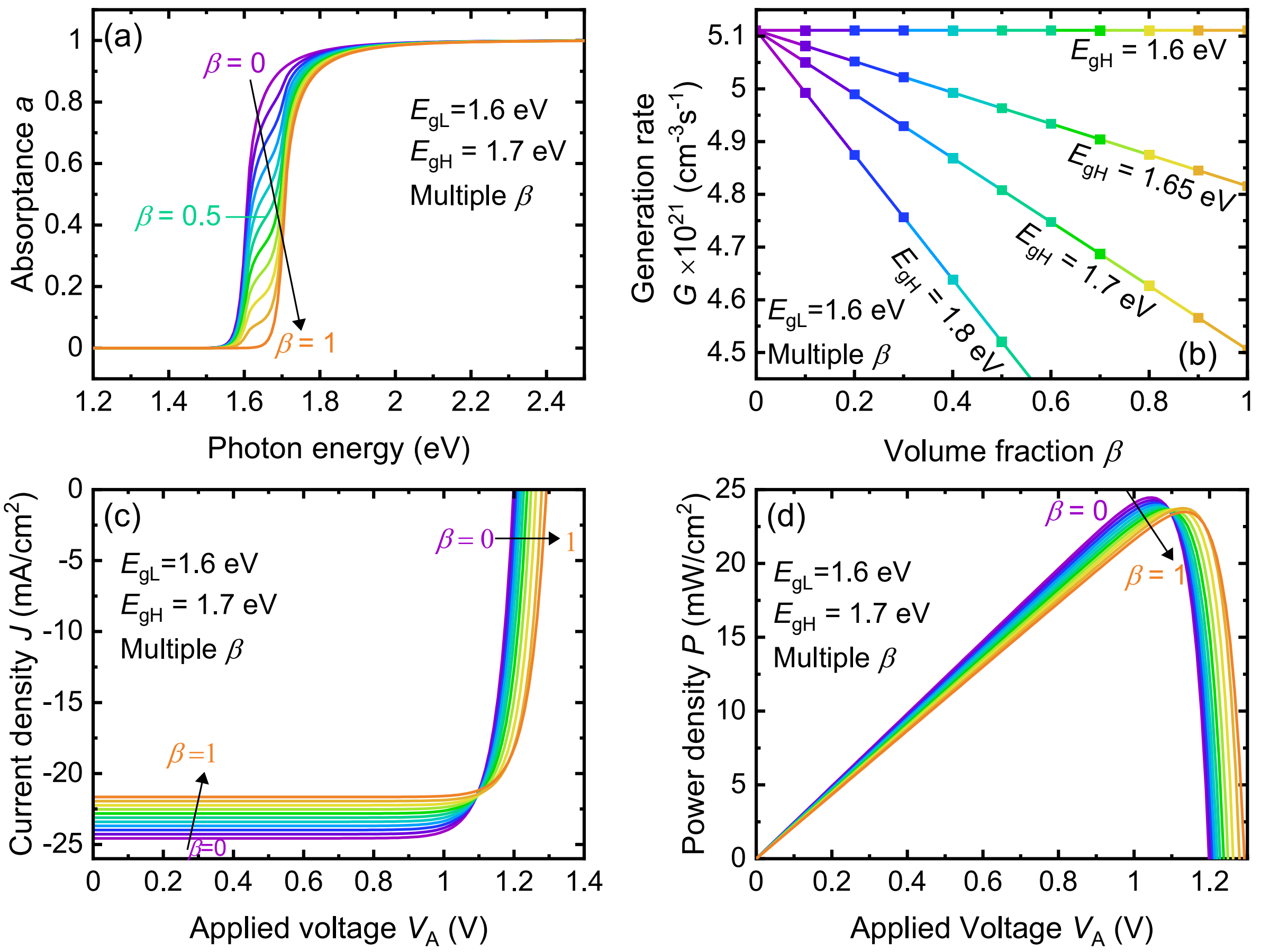

Figure S3. Effect of variation of bandgap and volume fraction $\beta$ on absorptance, Generation rate, current–voltage (*JV*) relationship, and power output. (a) Absorptance decreases for low energy photons as the volume fraction of the higher bandgap material $\beta$ increases. The absorptance also decreases for low energy photons as the higher bandgap increases. (b) The generation rate decreases both from increase in $\beta$ as well as $E_{gH}$ due to decrease in absorptance for low energy photons. (c) *JV* curves for varying values of $\beta$ when $E_{gL}$ =1.6 eV and $E_{gH}$ = 1.7 eV. The increase of the average band gap leads to a decrease of the equilibrium concentrations ($n_i^2$ goes down) and thereby to an increase of $np/n_i^2$ at open circuit, which leads to a higher open–circuit voltage. However, the short circuit current $J_{sc}$ decreases as we increase $\beta$ due to lower carrier generation. (d) The power density decreases as we increase $\beta$ due to decrease in $J_{sc}$.

# 3. Photodoping

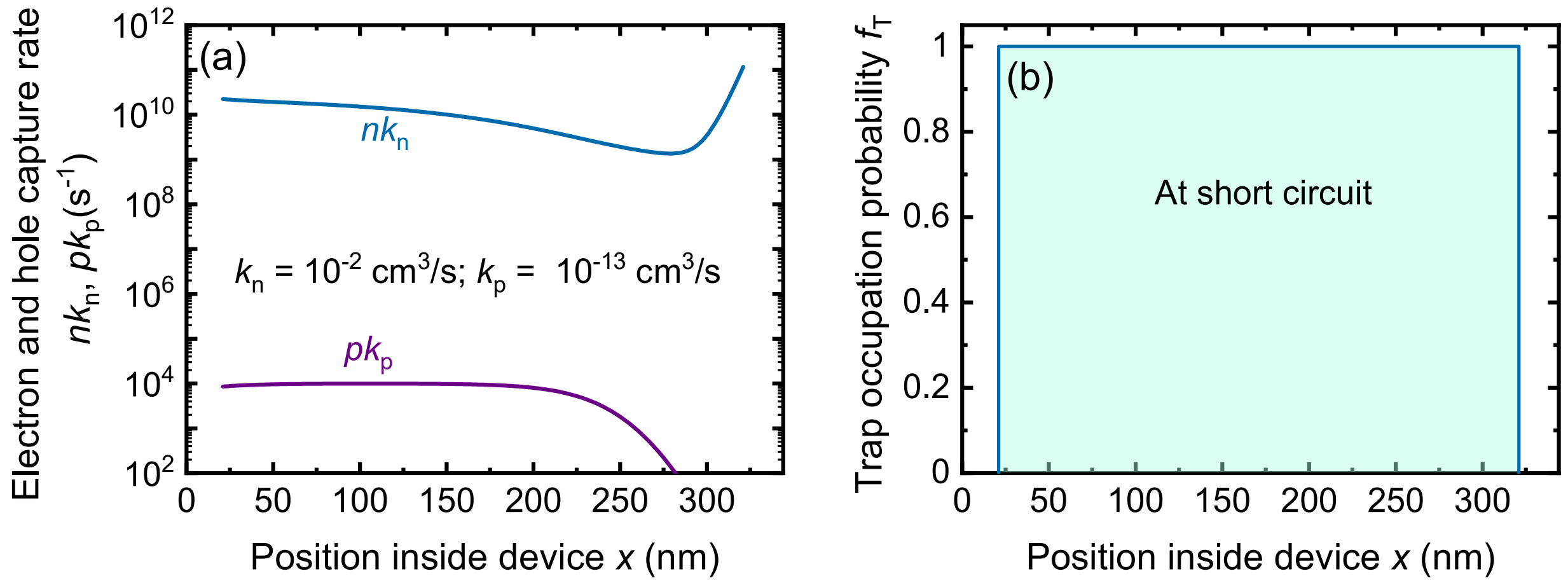

Figure S4. Photodoping due to presence of highly asymmetric capture coefficients of acceptor like defects. (a) Electron capture rate $nk_n$ is orders of magnitude higher than the hole capture rate $pk_p$. (b) When the defect level is within the two quasi–Fermi levels the electron emission rate and hole emission rate is very low. Hence, the only way a defect is emptied is by hole capture from valence band. Since the hole capture rate is orders of magnitude smaller than the electron capture rate, the defect stays filled by an electron and hence the trap occupation probability $f_T = 1$ throughout the device.

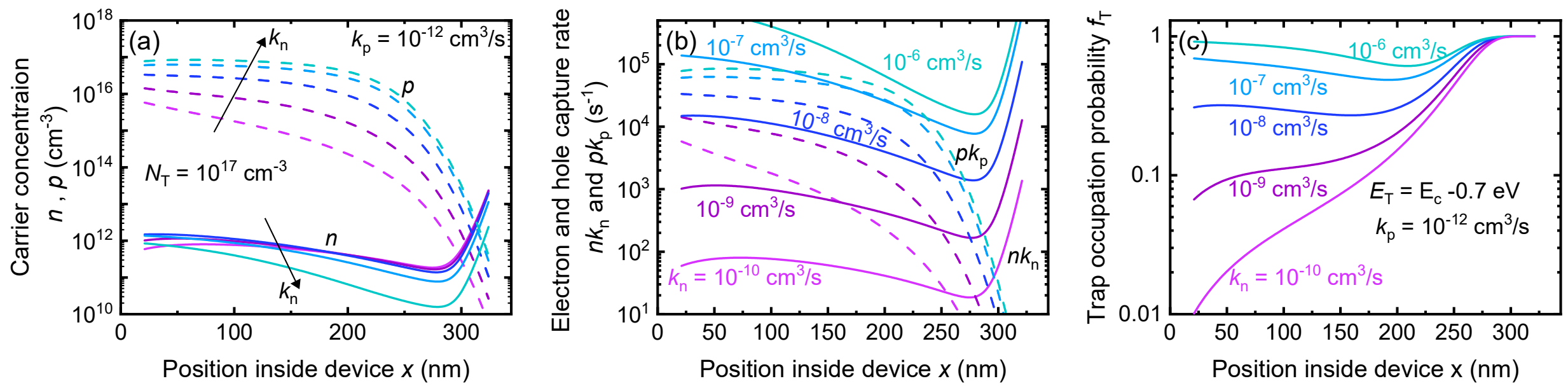

Figure S5. Photodoping due to presence of less asymmetric capture coefficients of acceptor type defects. (a) Electron and hole concentration inside the device due to various combinations of acceptor like defects. The hole concentration in the valence band increases with electron capture coefficient $k_n$, whereas the electron concentration $n$ decreases in the conduction band. This is because increased number of acceptor like defects trap electron from the conduction band and hence the occupation probability of the trap level increases as shown in panel (c). (b) Electron capture rate $nk_n$ (solid lines) for various values of $k_n$ is not always greater than hole capture rate for $k_p = 10^{-12}$ cm$^3$/s. The electron capture rate increases as we increase the value of $k_n$. However, the hole capture rate also increases due to increase in hole concentration in the valence band. (c) The trap occupation probability resulting from the respective values of $nk_n$ and $pk_p$. When $pk_p > nk_n$ , the occupation probability $f_T$ is lower than 1 but not zero. The trap occupation probability increases with the value of $nk_n$. As shown in Figure S4, the occupation probability would be $f_T = 1$ throughout the device only in the highly asymmetric case or when the $nk_n >> pk_p$. The relative values of the two capture rates will depend on the electron and hole concentration in the device, which subsequently depends on the electrostatic potential inside the device.

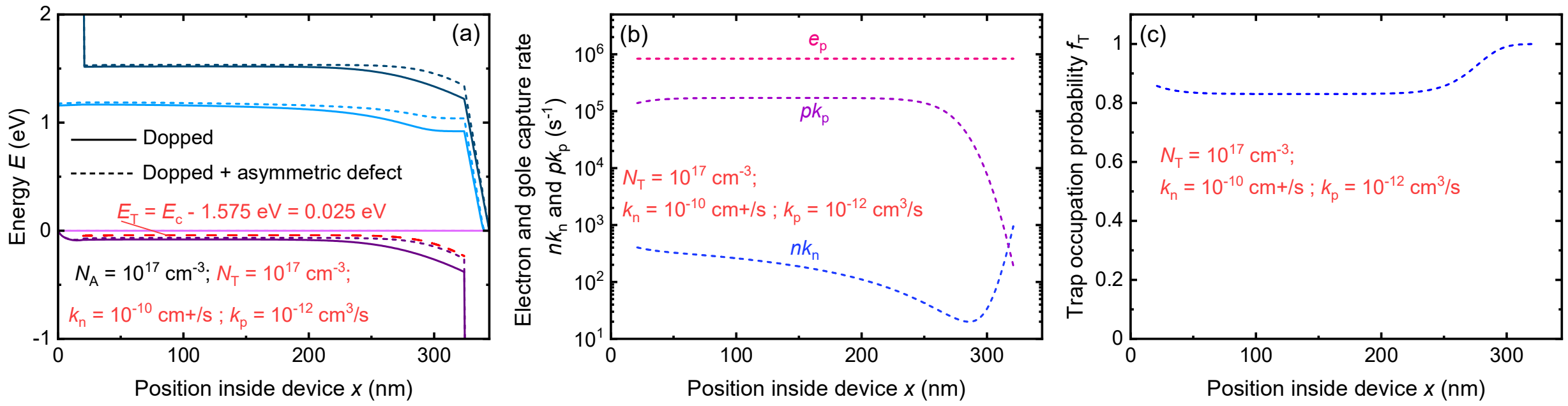

Figure S6: Band diagram , capture and emission rates, and occupation probability when an acceptor doped device has asymmetric defects present very close to the band edge.(a) (a) Band diagram of a device with an acceptor doping $N_A = 10^{17}$ cm$^{-3}$ and with and without an acceptor like asymmetric defect identified by $k_n = 10^{-10}$ cm$^3$/s and $k_p = 10^{-12}$ cm$^{-3}$ present 25 meV away from the valence band edge. The solid lines represent the doped device without the acceptor like defect level and the dashed line represent the device with an acceptor like defect level. The red dashed line very close the valence band represents the defect level 25 meV away from the valence band edge. (b) The capture and emission rates of the device with the acceptor like defect level. Since the defect is only 25 meV away from the valence band edge, the hole emission rate $e_p$, is higher than either of the two capture rates $nk_n$ and $pk_p$. The hole emission rate is responsible for filling the defect level with electrons by thermal emission. The electron emission rate is many orders of magnitude smaller than all the other rates and has no effect on the occupation probability of the defect (c) As a result of the very high hole emission rate $e_p$, the defect has a high probability of being occupied by an electron. However, it is important to note that the process filling the defect with electrons is thermally activated and is unlike the cases shown in previous figures where the defect is filled by electron capture from conduction band.

# 4. Material parameters

Table SII Material parameters for 0D model

| Parameter | Value |
|---|---|
| Low bandgap $E_{gL}$ | 1.6 eV |
| High bandgap $E_{gH}$ | 1.62, 1.65, 1.7, 1.8, 2.0 eV |
| Effective density of states $N_c = N_v$ | $2.2 \times 10^{18}$ cm$^{-3}$ |
| Thickness $x$ (Used in calculation of absorptance) | 300 nm |
| Urbach energy $U$ | 0.014 eV |
| Excess carrier concentration $\delta n$ | $10^{14} - 10^{18}$ cm$^{-3}$ |
| Mobility $\mu_{pero}$ (Used in calculation of diffusion length) | 30 cm$^2$/Vs |
| Defect capture coefficient $k_n = k_p$ | $10^{-8}$ cm$^3$/s |
| Auger coefficients $C_{aug} = c_n + c_p$ ($c_n = c_p$) | $10^{-28}$ cm$^6$/s |
| Radiative coefficients | Calculated from equation (S7) |
| Refractive index $\eta_r$ | 2.5 |

Table SIII Material parameters for device simulation using ASA

| | |
|---|---|
| Thickness of Absorber $d_{pero}$ | 300 nm |
| Thickness of Hole transport layer $d_{HTL}$ | 20 nm |
| Thickness of Electron transport layer $d_{ETL}$ | 20 nm |
| Thickness of interfaces (HTL/Pero, ETL/Pero) | 2 nm |
| Electron affinity of Absorber, $E_{A(Pero)}$ | 4 eV |
| Electron affinity of HTL, $E_{A(HTL)}$ | 2.6 eV |
| Electron affinity of ETL, $E_{A(ETL)}$ | 4 eV |
| Bandgap of Absorber, $E_{g(Pero)}$ | 1.6 eV |
| Bandgap of HTL, $E_{g(HTL)}$ | 3 eV |
| Bandgap of ETL, $E_{g(ETL)}$ | 3 eV |
| Bandgap of HTL/Pero interface | $E_{g(HTL)} + E_{A(HTL)} - E_{A(Pero)}$ |
| Bandgap of ETL/Pero interface | $E_{g(Pero)} + E_{A(Pero)} - E_{A(ETL)}$ |
| Mobility of Absorber, $\mu_{(Pero)}$[10] | 30 $cm^2/Vs$ |
| Mobility of HTL, ETL, $\mu_{(HTL,ETL)}$[11] | $10^{-2}$ ; $10^{-3}$; $10^{-4}$ $cm^2/Vs$ |
| Effective density of carriers (all layers) | $2.2 \times 10^{18}$ $cm^{-3}$ |
| Direct recombination coefficient (all layers) | $4.75 \times 10^{-10}$ $cm^3/s$ |
| Defect position from conduction band Ec – $E_T$ | 0.48 eV |
| Density of Donor traps: Absorber layer | $10^{15}$ $cm^{-3}$ |
| Interface | $10^{10}$ $cm^{-2}$ (figure 6–8)<br>$1.4 \times 10^{10}$ $cm^{-2}$ (figure 9) |
| Defect capture coefficient $k_n = k_p$ | $10^{-8}$ $cm^3/s$ (figure 6–8) |
| $k_n \neq k_p$ | $k_n = 0.7 \times 10^{-8}$ $cm^3/s$<br>$k_p = 0.2 \times 10^{-4}$ $cm^3/s$ (figure 9) |
| Auger coefficients $C_{aug} = c_n + c_p$ ($c_n = c_p$) | $10^{-28}$ $cm^6/s$ |
| Absorber and interface donor doping density $N_D$ | $10^{14} - 5\times10^{17}$ $cm^{-3}$ |
| Permittivity $\varepsilon$ | 10000 (figure 6–10) |

# 5. Generic device structure

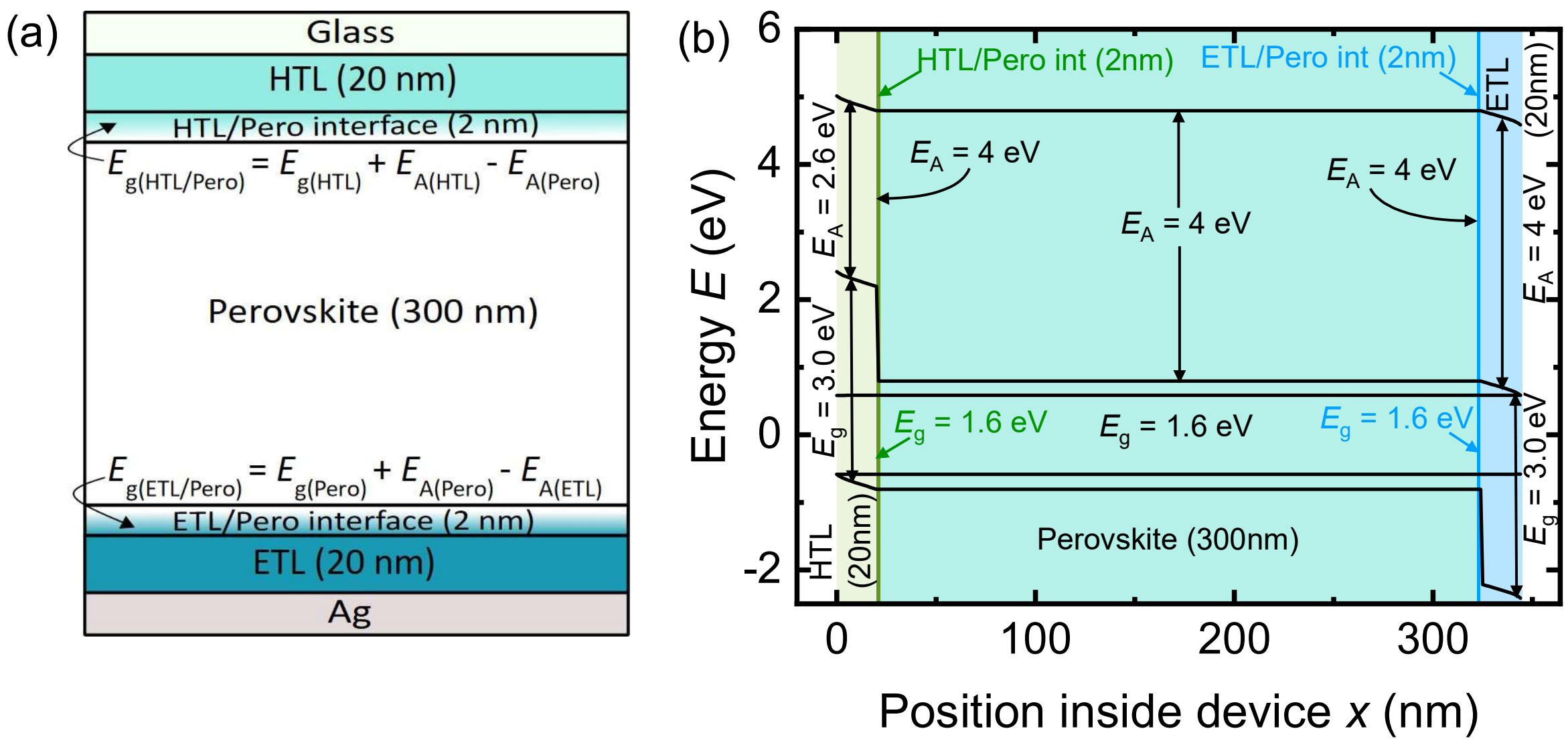

Figure S7. (a) A generic perovskite solar cell of p–i–n architecture used for all the analysis presented in the paper. (b) Band diagram of the generic device at open circuit condition.